\documentclass[10pt,journal,letterpaper,compsoc]{IEEEtran}
\newtheorem{theorem}{Theorem}

\makeatletter
\def\normalsize{\@setfontsize{\normalsize}{9.5bp}{12.00pt}}
\normalsize
\makeatother

\usepackage{graphics}
\usepackage{graphicx,color}
\usepackage{epsfig}
\usepackage{latexsym}
\usepackage{amsmath}
\usepackage{amssymb}
\usepackage{subfigure}
\usepackage{array}
\begin{document}
%
\title{Darknet-Based Inference of Internet Worm Temporal Characteristics}
%
%
%
%

\author{Qian~Wang,~\IEEEmembership{Student Member,~IEEE,}
        Zesheng~Chen,~\IEEEmembership{Member,~IEEE,}
         and~Chao~Chen,~\IEEEmembership{Member,~IEEE,}

\IEEEcompsocitemizethanks
{\IEEEcompsocthanksitem Q. Wang is
with the Department of Electrical and Computer Engineering,
Florida International University, Miami, FL, 33174.\protect\\
E-mail: qian.wang@fiu.edu.
\IEEEcompsocthanksitem Z. Chen and C. Chen are with the Department of
Engineering, Indiana University - Purdue University Fort Wayne, Fort Wayne, IN 46805.\protect\\
E-mail: \{zchen, chen\}@engr.ipfw.edu.}

}

\bibliographystyle{./IEEEtran}

\IEEEcompsoctitleabstractindextext{%
\begin{abstract}
Internet worm attacks pose a significant threat to network security
and management. In this work, we coin the term {\em Internet worm
tomography} as inferring the characteristics of Internet worms from
the observations of Darknet or network telescopes that monitor a
routable but unused IP address space. Under the framework of
Internet worm tomography, we attempt to infer Internet worm temporal
behaviors, {\em i.e.,} the host infection time and the worm
infection sequence, and thus pinpoint patient zero or initially
infected hosts. Specifically, we introduce statistical estimation
techniques and propose method of moments, maximum likelihood, and
linear regression estimators. We show analytically and empirically
that our proposed estimators can better infer worm temporal
characteristics than a naive estimator that has been used in the
previous work. We also demonstrate that our estimators can be
applied to worms using different scanning strategies such as random
scanning and localized scanning.

\end{abstract}

}


\maketitle

\IEEEdisplaynotcompsoctitleabstractindextext

%
\IEEEpeerreviewmaketitle

\section{Introduction} \label{intro}
%
%

%
%
%
%

\IEEEPARstart{S}{ince} Code Red and Nimda worms were released in
2001, epidemic-style attacks have caused severe damages. Internet
worms can spread so rapidly that existing defense systems cannot
respond until most vulnerable hosts have been infected. For example,
on January 25th, 2003, the Slammer worm reached its maximum scanning
rate of more than 55 million scans per second in about 3 minutes,
and infected more than 90\% of vulnerable machines within 10 minutes
\cite{Moore03}. It cost over one billion US dollars in cleanup and
economic damages. Therefore, worm attacks pose significant threats
to the Internet and meanwhile present tremendous challenges to the
research community.

To counteract these notorious plague-tide attacks, various detection
and defense strategies have been studied in recent years. According
to where the detectors are located, these strategies can generally
be classified into three categories: {\em source detection and
defenses}, detecting infected hosts in the local networks
\cite{Weaver04,Jung04,Williamson02,Khayam08}; {\em middle detection
and defenses}, revealing the appearance of worms by analyzing the
traffic going through routers \cite{Xie05,Chen04,Lakhina05}; and
{\em destination detection and defenses}, monitoring unwanted
traffic arriving at {\em Darknet or network telescopes}, a globally
routable address space where no active services or servers reside
\cite{Darknet,Telescope,Honeypots,IMS,Sink}. There are two types of
Darknet: {\em active Darknet} that responds to malicious scans to
elicit the payloads of the attacks \cite{Honeypots,IMS}, and {\em
passive Darknet} that observes unwanted traffic passively
\cite{Telescope,Sink}.

Different from source and middle detection and defenses, destination
detection and defenses offer unique advantages in observing
large-scale network explosive events such as distributed
denial-of-service (DDoS) attacks \cite{Moore06} and Internet worms
\cite{Moore02,Moore03,Shannon}. There is no legitimate reason for
packets destined to Darknet. Hence, most of the traffic arriving at
Darknet is malicious or unintended, including hostile reconnaissance
scans, probe activities from active worms, DDoS backscatter, and
packets from mis-configured hosts. Moreover, it has been shown that
for a large-scale worm event, most of infected hosts, if not all,
can be observed by the Darknet with a sufficiently large size \cite{AAWP}.

\begin{figure}[tb]
\includegraphics [width=0.5\textwidth, bb=92 316 613 627]{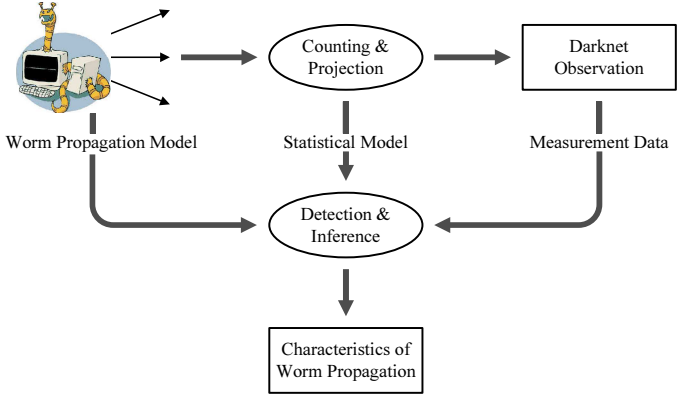}
\caption{Internet worm tomography.} \label{tomo}
\vspace{-0.3cm}
\end{figure}

In this work, we focus on the destination detection and defenses.
Specifically, we study the problem of inferring the characteristics
of Internet worms from Darknet observations. We refer to such a
problem as {\em Internet worm tomography}, as illustrated in
Fig.\ref{tomo}. Most worms use scan-based methods to find vulnerable
hosts and randomly generate target IP addresses. Thus, Darknet can
observe partial scans from infected hosts. Together with the worm
propagation model and the statistical model, Darknet observations
can be used to detect worm appearance
\cite{Wu04,Richardson05,Bu06,Soltani08} and infer worm
characteristics ({\em e.g.}, infection rate \cite{Zou05}, number of
infected hosts \cite{AAWP,Moore04}, and worm infection sequence
\cite{Kumar05,Rajab05,Wang08}). Internet worm tomography is named
after {\em network tomography}, which infers the characteristics of
the internal network ({\em e.g.}, link loss rate, link delay, and
topology) through the observations from end systems
\cite{Caceres,Coates}. Network tomography can be formulated as a
linear inverse problem. Internet worm tomography, however, cannot be
translated into the linear inverse problem due to the specific
properties of worm propagation, and thus presents new challenges.

Under the framework of Internet worm tomography, researchers have
studied worm temporal characteristics and have attempted to answer
the following important questions:
\begin{itemize}
\item {\em Host infection time:} When exactly does a specific host get
infected? This information is critical for the reconstruction of the worm infection
sequence \cite{Rajab05}.
\item{\em Worm infection sequence:} What is the order in which hosts are infected by worm
propagation? Such an order can help identify patient zero or initially infected
hosts \cite{Kumar05}.
\end{itemize}

The information of both the infection time and the infection sequence is important for defending
against worms. First, the identification of patient zero or
initially infected hosts and their infection times provide forensic
clues for law enforcement against the attackers who wrote and spread
the worm. Second, the knowledge of the infection sequence provides
insights into how a worm spread across the Internet ({\em e.g.,}
characteristics on who infected whom) and how network defense systems
were breached.


A simple estimator has been proposed in \cite{Rajab05}
to infer worm temporal behaviors. The estimator uses the observation
time when an infected host scans the Darknet for the first time as
the approximation of the host infection time to infer the worm
infection sequence. Such a naive estimator, however, does not fully
exploit all information obtained by the Darknet. Moreover, an
attacker can design a smart worm that uses lower scanning rates for
patient zero or initially infected hosts and higher scanning rates
for other infected hosts. In this way, the smart worm would weaken
the performance of the naive estimator.

The goal of this paper is to infer the Internet worm temporal
characteristics accurately by exploiting Darknet observations and
applying statistical estimation techniques. Our research work makes
several contributions:
\begin{itemize}
\item
We propose method of moments, maximum likelihood, and linear
regression statistical estimators to infer the host infection time.
We show analytically and empirically that the mean squared error of
our proposed estimators can be almost half of that of the naive
estimator in inferring the host infection time.
\item
We extend our proposed estimators to infer the worm infection
sequence. Specifically, we formulate the problem of estimating the
worm infection sequence as a detection problem and derive the
probability of error detection for different estimators. We
demonstrate analytically and empirically that our method performs
much better than the algorithm proposed in \cite{Rajab05}.
\item
We show empirically that our estimators have a better performance in
identifying patient zero or initially infected hosts of the smart
worm than the naive estimator. We also demonstrate that our
estimators can be applied to worms using different scanning
strategies such as random scanning and localized scanning.
\end{itemize}

The remainder of this paper is organized as follows. Section
\ref{time} introduces estimators for inferring the host infection
time. Section \ref{sequence} presents our algorithms in estimating
the worm infection sequence. Section \ref{simulation} gives
simulation results. Section \ref{discussion} discusses the
assumptions, the limitations, and the extensions of our estimators. Finally,
Section \ref{work} reviews related work, and Section \ref{conclude}
concludes the paper.

\section{Estimating the Host Infection Time} \label{time}

\begin{figure}[tb]
\includegraphics [width=0.5\textwidth, bb=89 366 541 567]{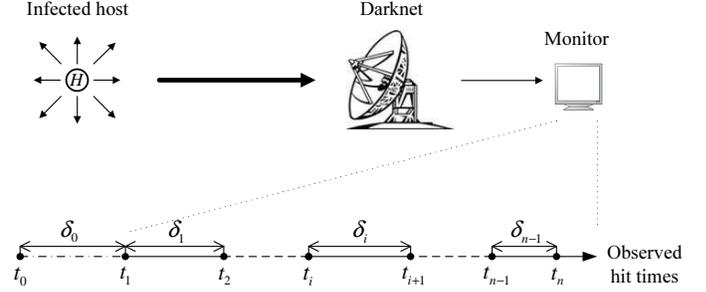}
\caption{An illustration of Darknet observations.} \label{darknet1}
\vspace{-0.3cm}
\end{figure}

We use Darknet observations to estimate when a host gets infected
and use {\it hit} to denote the event that a worm scan hits the
Darknet. As shown in Fig. \ref{darknet1}, suppose that a certain host is
infected at time $t_0$. The Darknet monitors a portion of the IPv4 address
space and can observe some scans from this host and record hit times
$t_1, t_2, \cdots, t_n$, where $n$ is the number of hit events from
this host. The problem of estimating the host infection time can
then be stated as follows: Given the Darknet observations $t_1, t_2,
\cdots, t_n$, what is the best estimate of $t_0$?

To study this problem, we make the following assumptions: 1) There
is no packet loss in the Internet. 2) An infected host uses its
actual source IP address and does not apply IP spoofing, which is
the case for TCP worms. 3) The scanning rate $s$ ({\it i.e.}, the
number of scans sent by an infected host per time unit) is
time-invariant for an infected host, whereas the scanning rates of infected hosts can be different from each other.
The last assumption comes from the observation that famous worms, such as Code Red, Nimda, Slammer,
and Witty, do not apply any scanning rate variation mechanisms.
An infected host always scans for vulnerable hosts at the maximum speed allowed by its computing resources and network conditions \cite{Wei}.
In Section \ref{discussion}, we will revisit and discuss these assumptions.

Obviously, inferring $t_0$ from Darknet observations is affected by
the Internet-worm scanning methods. In this paper, we focus on
random scanning and localized scanning. However, if a scan from an
infected host hits Darknet with a time-invariant probability, our
estimation techniques are independent of worm-scanning methods.
To analytically estimate the host infection time, we consider a discrete-time system. For random
scanning (RS), a worm selects targets randomly and scans the entire
IPv4 address space with $\Omega$ addresses ({\em i.e.,}
$\Omega=2^{32}$). We assume that Darknet monitors $\omega$
addresses. Thus, the probability for a scan to hit the Darknet is
$\omega/\Omega$; and the probability of a hit event in the
discrete-time system ({\it i.e.}, the probability that Darknet
observes at least one scan from the same infected host in a time
unit) is
    \begin{equation}
    \label{equ:p_rs}
         \mbox{Pr}_{\mbox {\tiny RS}} (\mbox{hit event}) = 1-\Big(1- \frac{\omega}{\Omega}\Big)^s.
    \end{equation}
Since $s$ is time-invariant for a given infected host,
$\mbox{Pr}_{\mbox {\tiny RS}} (\mbox{hit event})$ is also
time-invariant.

Localized scanning (LS) preferentially searches for vulnerable hosts
in the ``local" address space \cite{Chen07}. For simplicity, in this
paper we only consider the $/l$ LS: $p_a (0 \leq p_a < 1)$ of the
time, a ``local" address with the same first $l$ bits as the
attacking host is chosen as the target; $1- p_a$ of the time, a
random address is chosen. We consider a centralized Darknet that
occupies a continuous address space and monitors $\omega$ addresses.
Moreover, we assume that the Darknet is contained in a $/l$ prefix with no vulnerable hosts. For example, network telescopes
used by CAIDA are such a centralized Darknet and contain a $/8$
subnet. Since no infected hosts exist in the $/l$ subnet where the Darknet resides,
the probability for a worm scan to hit the
Darknet is $(1-p_a)\cdot \omega/\Omega$. Therefore, the probability
of a hit event in the discrete-time system is
    \begin{equation}
    \label{equ:p_ls}
         \mbox{Pr}_{\mbox {\tiny LS}} (\mbox{hit event}) = 1-\Big(1-
         (1-p_a)\cdot\frac{\omega}{\Omega}\Big)^s,
    \end{equation}
which is time-invariant. Since $\mbox{Pr}_{\mbox {\tiny RS}}
(\mbox{hit event})$ has a similar form as $\mbox{Pr}_{\mbox {\tiny
LS}} (\mbox{hit event})$ and is the special case of
$\mbox{Pr}_{\mbox {\tiny LS}} (\mbox{hit event})$ when $p_a$ = 0, we
use $p$ ($0<p<1$) to denote the hit probability in general for both cases to
simplify our discussion.

\begin{table}[tb]
    \caption{Notations used in this paper.}
    \def\temptablewidth{0.5\textwidth}
    {\rule{\temptablewidth}{1pt}}
\begin{tabular*}{\temptablewidth}{@{\extracolsep{\fill}}cp{7cm}}
Notations & \hspace*{2.5cm} Definition \\
\hline
$\Omega$    & Size of the scanning space ($\Omega=2^{32}$)\\
$\omega$    & Size of the Darknet\\
$s$         & Scanning rate (scans/time unit)\\
$\sigma$    & Standard deviation of the scanning rate \\
$p_a$       & Probability that an address with the same first $l$ bits as the attacking host is chosen by LS \\
$p$         & Probability that at least one scan from the same infected host hits the Darknet in a time unit\\
\hline
$t_0$       & Host infection time \\
$\hat{t}_0$ & Estimated host infection time \\
$t_i$       & Discrete time tick when the infected host hits the Darknet for the $i$-th time $(i\geq 1)$ \\
$\delta_i$  & Time interval between two consecutive hits of the Darknet ($\delta_i = t_{i+1} - t_i$, $i\geq 1$)\\
$n$         & Number of hit events observed at the Darknet for an infected host\\
$\mu$       & Mean of $\delta$ \\
$\hat{\mu}$ & Estimation of $\mu$ \\
\hline
$D$         & Sequence distance \\
$S_i$       & Worm infection sequence \\
$\hat{S_i}$ & Estimated worm infection sequence \\
$N$         & Length of the worm infection sequence considered for evaluation\\
\end{tabular*}
    {\rule{\temptablewidth}{1pt}}
    \label{tab:notations}
        \vspace{-0.3cm}
\end{table}

Denote $\delta_0$ as the time interval between when a host gets
infected and when Darknet observes the first scan from this host,
{\em i.e.,} $\delta_0=t_1-t_0$, as shown in Fig. \ref{darknet1}.
Denote $\delta_i$ as the time interval between $i$-th hit and
$(i+1)$-th hit on Darknet, {\em i.e.,} $\delta_i=t_{i+1}-t_i$, $i\ge
1$. Thus, $\delta_0, \delta_1, \cdots, \delta_{n-1}$ are independent
and identically distributed (i.i.d.) and follow a geometric
distribution with parameter $p$, {\em i.e.,}
    \begin{equation}
         \mbox{Pr}(\delta=k) = p \cdot (1-p)^{k-1},\ k=1,2,3,\cdots ,
         \label{equ:geo}
    \end{equation}
    \begin{equation}
        \mbox{E}(\delta) = {1\over p}= \mu, \ \ \ \ \   \mbox{Var}(\delta) = \frac{1-p}{p^2}.
    \end{equation}
Denote $\mu$ as the mean value of $\delta$ and $\hat{\mu}$ as the estimate of $\mu$.
We then estimate $t_0$ by subtracting $\hat{\mu}$ from $t_1$, {\em i.e.,}
    \begin{equation}
          \hat{t_0} = t_1 - \hat{\mu}.
    \end{equation}
Therefore, our problem is reduced to estimating $\mu$. Table
\ref{tab:notations} summarizes the notations used in this paper.

\subsection{Naive Estimator}
Since $\delta$ follows the geometric distribution as described by
Equation (\ref{equ:geo}), $\mbox{Pr}(\delta)$ is maximized when
$\delta =1$. Then, a {\em naive estimator} (NE) of $\mu$ is
    \begin{equation}
         \hat{\mu}_{\mbox {\tiny NE}} = 1.
    \end{equation}
Thus, the NE of $t_0$ is
    \begin{equation}
        \hat{t_0}_{\mbox {\tiny NE}} = t_1 - \hat{\mu}_{\mbox {\tiny NE}} = t_1 - 1.
    \end{equation}
Note that $\hat{t_0}_{\mbox {\tiny NE}}$ depends only on $t_1$, but
not on $t_2, t_3, \cdots, t_n$. This estimator has been used in
\cite{Rajab05} to infer the host infection time and the worm
infection sequence. In this paper, however, we consider more
advanced estimation methods.

\subsection{Method of Moments Estimator}
Since $\mbox{E}(\delta) = \mu$, we design a {\em method of moments
estimator} (MME), {\em i.e.,}
    \begin{equation}
    \label{equ:MME_u}
        \hat{\mu}_{\mbox {\tiny MME}} = \overline \delta = {1 \over n-1}{\sum\limits_{i = 1}^{n - 1}{\delta_i}}= \frac{t_n-t_1}{n-1}.
    \end{equation}
Thus, the MME of $t_0$ is
    \begin{equation}
          \hat{t_0}_{\mbox {\tiny MME}} = t_1 - \hat{\mu}_{\mbox {\tiny MME}} = t_1 - \frac{t_n-t_1}{n-1}.
    \end{equation}
Note that $\hat{t_0}_{\mbox {\tiny MME}}$ is not only related to
$t_1$, but also to $n$ and $t_n$.

\subsection{Maximum Likelihood Estimator}
Rewrite the probability mass function of $\delta$ in Equation
(\ref{equ:geo}) with respect to $\mu$,
    \begin{equation}
        \textstyle \mbox{Pr}(\delta;\mu) = {1\over \mu}\Big(1-{1\over \mu}\Big)^{\delta-1}, \delta=1,2,3,\cdots.
    \end{equation}
Since $\delta_1,\delta_2, \cdots, \delta_{n-1}$ are i.i.d., the likelihood function is
given by the following product
    \begin{eqnarray}
         \mbox{L}(\mu) & =&  \prod\limits_{i = 1}^{n - 1} \mbox{Pr}(\delta_i; \mu)  \nonumber\\
         & = &   \Big({1\over \mu}\Big)^{n-1} \Big(1-{1\over \mu} \Big)^{ (\sum\limits_{i = 1}^{n- 1} {\delta_i})-(n-1)}.
    \end{eqnarray}
We then design a {\em maximum likelihood estimator} (MLE), {\em i.e.,}
    \begin{equation}
         \hat{\mu}_{\mbox {\tiny MLE}} =  \arg\max_{\mu} \mbox{L}(\mu).
    \end{equation}
Rather than maximizing $\mbox{L}(\mu)$, we choose to maximize its logarithm $ln\mbox{L}(\mu)$. That is,
   \begin{equation}
        \frac{d}{d\mu}{ln\mbox{L}(\mu)} = 0
   \end{equation}
      \begin{equation}
      \label{equ:MLE_u}
         \Longrightarrow \hat{\mu}_{\mbox {\tiny MLE}}  = {1 \over n-1}{\sum\limits_{i = 1}^{n - 1} {\delta_i}}=\frac{t_n-t_1}{n-1},
   \end{equation}
which has the same expression as the MME. Thus,
    \begin{equation}
        \hat{t_0}_{\mbox {\tiny MLE}} = t_1 - \hat{\mu}_{\mbox {\tiny MLE}} = t_1 - \frac{t_n-t_1}{n-1}.
    \end{equation}

\subsection{Linear Regression Estimator}
\begin{figure}[tb]
\begin{center}
   \includegraphics [width=2.4in, bb=196 176 501 442]{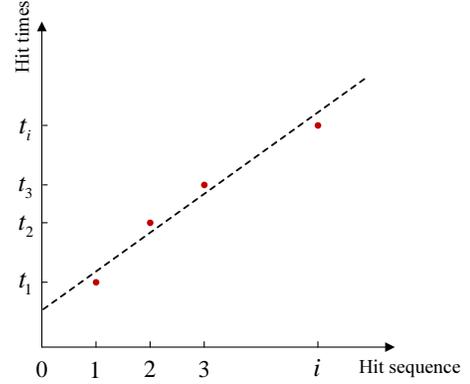}
\caption{Linear
regression model.} \label{regression}
\end{center}
\end{figure}

Under the assumption that the scanning rate of an individual
infected host is time-invariant, the relationship between $t_i$ and
$i$ can be described by a linear regression model as illustrated in
Fig. \ref{regression}, {\em i.e.,}
    \begin{equation}
         t_i = \alpha + \beta \cdot i + \varepsilon_i,
    \end{equation}
where $\alpha$ and $\beta$ are coefficients, and $\varepsilon_i$ is the error term.
To fit the observation data, we apply the least squares method to adjust the parameters of the model.
That is, we choose the coefficients that minimize the residual sum of squares (RSS)
    \begin{equation}
        \mbox {RSS} = \sum\limits_{i = 1}^n {[t_i  - (\alpha  + \beta \cdot i)]^2}.
\end{equation}
The minimum RSS occurs when the partial derivatives with respect to the coefficients are zero
    \begin{equation}
        \left\{ \begin{aligned} \frac{{\partial  \mbox {RSS}}}{{\partial \alpha }} &
        =-2\sum\limits_{i = 1}^n {(t_i  - \alpha  - \beta \cdot i) = 0}\\
        \frac{{\partial  \mbox {RSS}}}{{\partial \beta }} &=  -
        2\sum\limits_{i = 1}^n {i \cdot  (t_i  - \alpha  - \beta \cdot i)  =
        0},
        \end{aligned} \right.
    \label{rss}
    \end{equation}
which leads to
    \begin{equation}
        \left\{ \begin{aligned} \hat{\alpha} & = \overline  t - \hat{\beta} \cdot \overline i\\
        \hat{\beta} &= \frac{\overline {i \cdot t}- \overline i \cdot
        \overline t}{\overline {i^2} - (\overline i) ^2},
        \end{aligned} \right.
    \end{equation}
\begin{table}[!h]
   \caption{Comparison of estimator properties ($\hat{\mu}$).}
   \def\temptablewidth{0.5\textwidth}
{\rule{\temptablewidth}{1pt}}
\begin{tabular*}{\temptablewidth}{@{\extracolsep{\fill}}llll}
$\hat{\mu}$ & $\mbox{Bias}(\hat{\mu}) $ &  $\mbox{Var}(\hat{\mu})$ & $\mbox{MSE}(\hat{\mu})$ \\
\hline $\hat{\mu}_{\mbox {\tiny NE}} = 1$ & $1 - {1\over p}$ & $0$ & $\frac{(1-p)^2}{p^2}$  \\
$\hat{\mu}_{\mbox {\tiny MME}} =\hat{\mu}_{\mbox {\tiny MLE}}= \frac{t_n-t_1}{n-1}$ & $0$ & $\frac{1-p}{p^2(n-1)}$ &  $\frac{1-p}{p^2(n-1)}$ \\
$\hat{\mu}_{\mbox {\tiny LRE}} = \frac{\overline {i \cdot t}-
\overline i \cdot
\overline t}{\overline {i^2} - (\overline i) ^2}$ & $0$ &$\frac{6(n^2+1)(1-p)}{5n(n^2-1)p^2}$ &$\frac{6(n^2+1)(1-p)}{5n(n^2-1)p^2}$\\
\end{tabular*}
       {\rule{\temptablewidth}{1pt}}
           \label{tab:mu}
           \vspace{-0.2cm}
\end{table}
where the bar symbols denote the average values
    \begin{equation}
    \label{equ:LRE_1}
        \left\{ \begin{aligned} \overline i &= \frac{1}{n}\sum\limits_{i =
        1}^n i,\ \ \ \overline {i^2} = \frac{1}{n}\sum\limits_{i = 1}^n i^2 \\
        \overline {t} &= \frac{1}{n}\sum\limits_{i = 1}^n t_i,\ \
        \overline {i\cdot t} = \frac{1}{n}\sum\limits_{i = 1}^n i \cdot t_i.
        \end{aligned} \right.
    \end{equation}
We then design a {\em linear regression estimator} (LRE), {\em i.e.,}
\begin{equation}
        \hat{\mu}_{\mbox {\tiny LRE}} = \hat{\beta} =\hat{t_1} - \hat{t_0}.
    \end{equation}
Thus, the LRE of $t_0$ is
    \begin{equation}
        \hat{t_0}_{\mbox {\tiny LRE}} =t_1 - \hat{\mu}_{\mbox {\tiny LRE}} = t_1 - \frac{\overline {i \cdot t}- \overline i \cdot
        \overline t}{\overline {i^2} - (\overline i) ^2}.
    \end{equation}

There is another way to estimate $t_0$, which uses the point of interception shown in Fig. \ref{regression} as the
estimation of $t_0$, {\em i.e.,}
    \begin{equation}
        \hat{t_0}^{'}_{\mbox {\tiny LRE}} = \hat{\alpha} = \overline  t - \hat{\mu}_{\mbox {\tiny LRE}} \cdot \overline
        i.
    \end{equation}
However, we find that the mean squared error of $\hat{t_0}^{'}_{\mbox
{\tiny LRE}}$ increases when $n$ increases.
That is, the performance of the estimator worsens with the increasing number of hits, which makes this estimator undesirable.

\subsection{Comparison of Estimators}
\begin{table*}
   \caption{Comparison of estimator properties ($\hat{t_0}$).}
   \def\temptablewidth{\textwidth}
{\rule{\temptablewidth}{1pt}}
\begin{tabular*}{\temptablewidth}{@{\extracolsep{\fill}}lcll}
$\hat{t_0}$ & $\mbox{Bias}(\hat{t_0}) $ &  $\mbox{Var}(\hat{t_0})$ & $\mbox{MSE}(\hat{t_0})$\\
\hline $\hat{t_0}_{\mbox {\tiny NE}} = t_1 - \hat{\mu}_{\mbox {\tiny
NE}}$ & $\frac{1-p}{p}$ & $\frac{1-p}{p^2}$ &
$\frac{(1-p)(2-p)}{p^2} \ \ \ \ \ \ \ \ \ \ \ \ \ \ \ (\approx \frac{2(1-p)}{p^2},\ \mbox{when}\ p\ll 1)$\\
$\hat{t_0}_{\mbox {\tiny MME}}=\hat{t_0}_{\mbox {\tiny MLE}} = t_1 - \hat{\mu}_{\mbox {\tiny
MME}}$ & $0$ & $  \frac{1-p}{p^2} \cdot \frac {n}{n-1}$  &
$ \frac{1-p}{p^2} \cdot \frac {n}{n-1} \ \ \ \ \ \ \ \ \ \ \ \ \ \ \ \ (\approx \frac{1-p}{p^2},\ \mbox{when}\ n\gg 1)$ \\
$\hat{t_0}_{\mbox {\tiny LRE}} = t_1 -  \hat{\mu}_{\mbox {\tiny
LRE}} $ & $0$ & $\frac{1-p}{p^2} \cdot
\frac{5n^3+6n^2-5n+6}{5n(n^2-1)}$& $\frac{1-p}{p^2} \cdot
\frac{5n^3+6n^2-5n+6}{5n(n^2-1)} \ \ (\approx \frac{1-p}{p^2},\ \mbox{when}\ n\gg 1)$\\
\end{tabular*}
       {\rule{\temptablewidth}{1pt}}
     \label{tab:t0}
     \vspace{-0.4cm}
\end{table*}
To compare the performance of the naive estimator and our proposed
estimators, we compute the bias, the variance, and the mean squared
error (MSE). For estimating $\mu$,
    \begin{equation}
        \left\{ \begin{array}{l l}  \mbox{Bias}(\hat{\mu}) \!\!\!\! &= \,  \mbox{E}(\hat{\mu})-\mu \\
         \mbox{Var}(\hat{\mu}) \!\!\!\! & =  \, \mbox{E}\,[(\hat{\mu}-\mbox{E}(\hat{\mu}))^2]\\
    \mbox{MSE}(\hat{\mu}) \!\!\! \! &  = \, \mbox{E}\,[(\hat{\mu}-\mu)^2]=\mbox{Bias}^2(\hat{\mu})+\mbox{Var}(\hat{\mu}).\\
        \end{array} \right.
    \end{equation}
Here, the {\em bias} denotes the average deviation of the estimator
from the true value; the {\em variance} indicates the distance
between the estimator and its mean; and the {\em MSE} characterizes
the closeness of the estimated value to the true value. A smaller MSE
indicates a better estimator. Table \ref{tab:mu} summarizes the
results of NE, MME (or MLE), and LRE for estimating $\mu$. The
details of the derivations of Table \ref{tab:mu} are given in
Appendix A. It is noted that MME and LRE are unbiased, while NE is
biased. Moreover, MME and LRE have a smaller MSE than NE if $n>$ 2
and $p<$ 0.5, a condition that is usually satisfied. Specifically,
when $n\rightarrow \infty$, $\mbox{MSE}(\hat{\mu}_{\mbox {\tiny
MME}})\rightarrow 0$ and $\mbox{MSE}(\hat{\mu}_{\mbox {\tiny
LRE}})\rightarrow 0$, but $\mbox{MSE}(\hat{\mu}_{\mbox {\tiny
NE}})\rightarrow (1-p)^2/p^2$. It is also observed that MME is slightly
better than LRE in terms of MSE when $n>$ 2.

Similarly, we compute the bias, the variance, and the MSE of the
estimators for estimating $t_0$ in Table \ref{tab:t0}. The details
of the derivations of Table \ref{tab:t0} are given in Appendix B. We
also observe that MME (or MLE) and LRE are unbiased, whereas NE is
biased. Moreover, $\mbox{MSE}(\hat{t_0}_{\mbox {\tiny MME}})$ and
$\mbox{MSE}(\hat{t_0}_{\mbox {\tiny LRE}})$ are smaller than
$\mbox{MSE}(\hat{t_0}_{\mbox {\tiny NE}})$, and
$\mbox{MSE}(\hat{t_0}_{\mbox {\tiny MME}})$ is the smallest when
$n>$ 3 and $p<$ 0.5. Specifically, in practice, Darknet only covers a
relatively small portion of the IPv4 address space ({\em i.e.,}
$\omega \ll \Omega$), which leads to $p \ll 1$. Thus, we have the
following theorem:
\begin{theorem}
\label{thm:time}
When the Darknet observes a sufficient number of
hits ({\em i.e.,} $n \gg 1$) and $p \ll 1$,
\begin{equation}
\mbox{MSE}(\hat{t_0}_{\mbox {\tiny MME}})\approx
\mbox{MSE}(\hat{t_0}_{\mbox {\tiny LRE}})\approx
\frac{1}{2}\mbox{MSE}(\hat{t_0}_{\mbox {\tiny NE}}).
\end{equation}
\end{theorem}
That is, the MSE of our proposed estimators is almost half of
that of the naive estimator. That is, our proposed estimators are nearly twice as accurate as the naive estimator in estimating the host infection time.

\section{Estimating the Worm Infection Sequence} \label{sequence}

In this section, we extend our proposed estimators for inferring the
worm infection sequence.

\subsection{Algorithm}

Our algorithm is that we first estimate the infection time of each
infected host. Then, we reconstruct the infection sequence based on
these infection times. That is, if $\hat{t_{\mbox{\tiny
0}}}_{\mbox{\tiny A}} < \hat{t_{\mbox{\tiny 0}}}_{\mbox {\tiny B}}$,
we infer that host A is infected before host B. It is noted that the
algorithm used in \cite{Rajab05} to infer the worm infection
sequence can be regarded as using this approach with the naive
estimator.

\begin{figure}[tb]
\begin{center}
\includegraphics [width=0.5\textwidth, bb=89 327 543 423]{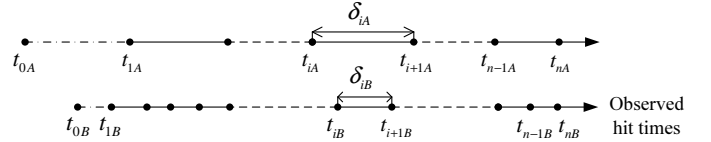}
\caption{A scenario of the worm infection sequence.} \label{darknet2}
\end{center}
\vspace{-0.3cm}
\end{figure}

The naive estimator, however, can potentially fail to infer the worm
infection sequence in some cases. Fig. \ref{darknet2} shows an
example, where hosts A and B get infected at $t_{\mbox{\tiny 0A}}$
and $t_{\mbox {\tiny 0B}}$, respectively, and $t_{\mbox{\tiny 0A}} <
t_{\mbox {\tiny 0B}}$. Moreover, these two infected hosts have
scanning rates $s_{\mbox{\tiny A}}<s_{\mbox{\tiny B}}$ such that Darknet
observes $t_{\mbox{\tiny 1A}} > t_{\mbox{\tiny 1B}}$. If the naive
estimator is used, $\hat{t_{\mbox{\tiny 0}}}_{\mbox{\tiny A}} >
\hat{t_{\mbox{\tiny 0}}}_{\mbox {\tiny B}}$, which means that host A
is incorrectly inferred to be infected after host B. Intuitively, if
our proposed estimators are applied, it is possible to obtain
$\hat{t_{\mbox{\tiny 0}}}_{\mbox{\tiny A}} < \hat{t_{\mbox{\tiny
0}}}_{\mbox {\tiny B}}$ and thus recover the real infection sequence.

\subsection{Performance Analysis}

To analytically show that our estimators are more accurate than the
naive estimator in estimating the worm infection sequence, we
formulate the problem as a detection problem. Specifically, in Fig.
\ref{darknet2}, suppose that host B is infected after host A ({\em
i.e.,} $t_{\mbox{\tiny 0A}} < t_{\mbox {\tiny 0B}}$). If
$\hat{t_{\mbox{\tiny 0}}}_{\mbox{\tiny A}} < \hat{t_{\mbox{\tiny
0}}}_{\mbox {\tiny B}}$, we call it ``success" detection; otherwise,
if $\hat{t_{\mbox{\tiny 0}}}_{\mbox{\tiny A}} > \hat{t_{\mbox{\tiny
0}}}_{\mbox {\tiny B}}$, we call it ``error"
detection\footnote{We ignore the case $\hat{t_{\mbox{\tiny
0}}}_{\mbox{\tiny A}} = \hat{t_{\mbox{\tiny 0}}}_{\mbox {\tiny B}}$
here.}. We intend to calculate the probability of error detection
for different estimators.

Note that $\delta_{\mbox{\tiny 0A}}=t_{\mbox{\tiny
1A}}-t_{\mbox{\tiny 0A}}$ and $\delta_{\mbox{\tiny
0B}}=t_{\mbox{\tiny 1B}}-t_{\mbox{\tiny 0B}}$ follow the geometric
distribution ({\em i.e.,} Equation (\ref{equ:geo})) with parameter
$p_{\mbox{\tiny A}}$ and $p_{\mbox{\tiny B}}$, respectively. Here,
$p_{\mbox{\tiny A}}$ (or $p_{\mbox{\tiny B}}$) is the probability
that at least one scan from host A (or B) hits the Darknet in a time
unit and follows Equation (\ref{equ:p_rs}) for random scanning and
Equation (\ref{equ:p_ls}) for localized scanning. Moreover,
$p_{\mbox{\tiny A}}$ (or $p_{\mbox{\tiny B}}$) depends on
$s_{\mbox{\tiny A}}$ (or $s_{\mbox{\tiny B}}$) so that if
$s_{\mbox{\tiny A}} < s_{\mbox{\tiny B}}$, then $p_{\mbox{\tiny A}}
< p_{\mbox{\tiny B}}$. Since $\omega \ll \Omega$, we have
$p_{\mbox{\tiny A}} \ll 1$ and $p_{\mbox{\tiny B}} \ll 1$. Hence,
for simplicity we use the continuous-time analysis and apply the
exponential distribution to approximate the geometric distribution
for $\delta_{\mbox{\tiny 0A}}$ and $\delta_{\mbox{\tiny 0B}}$
\cite{Jain}, {\em i.e.,}
\begin{equation}
f(x;\lambda)=
\left\{\begin{array} {l l} \lambda e^{-\lambda x}, & \quad  x \geq 0\\
0, & \quad  x< 0,\\
\end{array} \right.
\label{eqn:exp}
\end{equation}
where $\lambda=p_{\mbox{\tiny A}}$ or $p_{\mbox{\tiny B}}$.

\begin{figure*}[tb]
\begin{center}
    \mbox{
      \subfigure[$\mbox{Pr}_{\mbox {\tiny NE}}(\mbox{error})$.]{\includegraphics[width=2.7in]{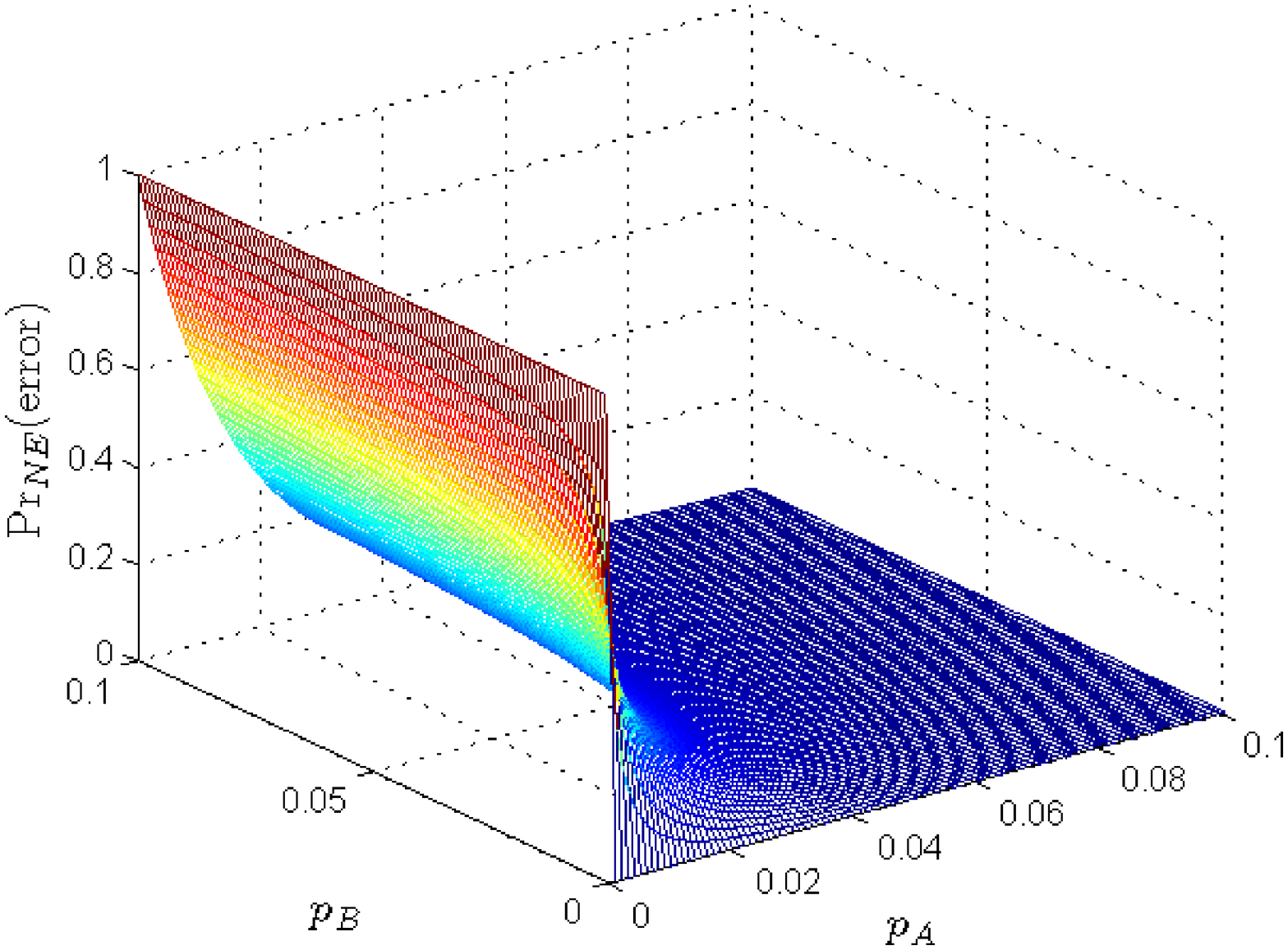}}
      \hspace{2cm}
       \subfigure[$\mbox{Pr}_{\mbox {\tiny MME}}(\mbox{error})$.]{\includegraphics[width=2.7in]{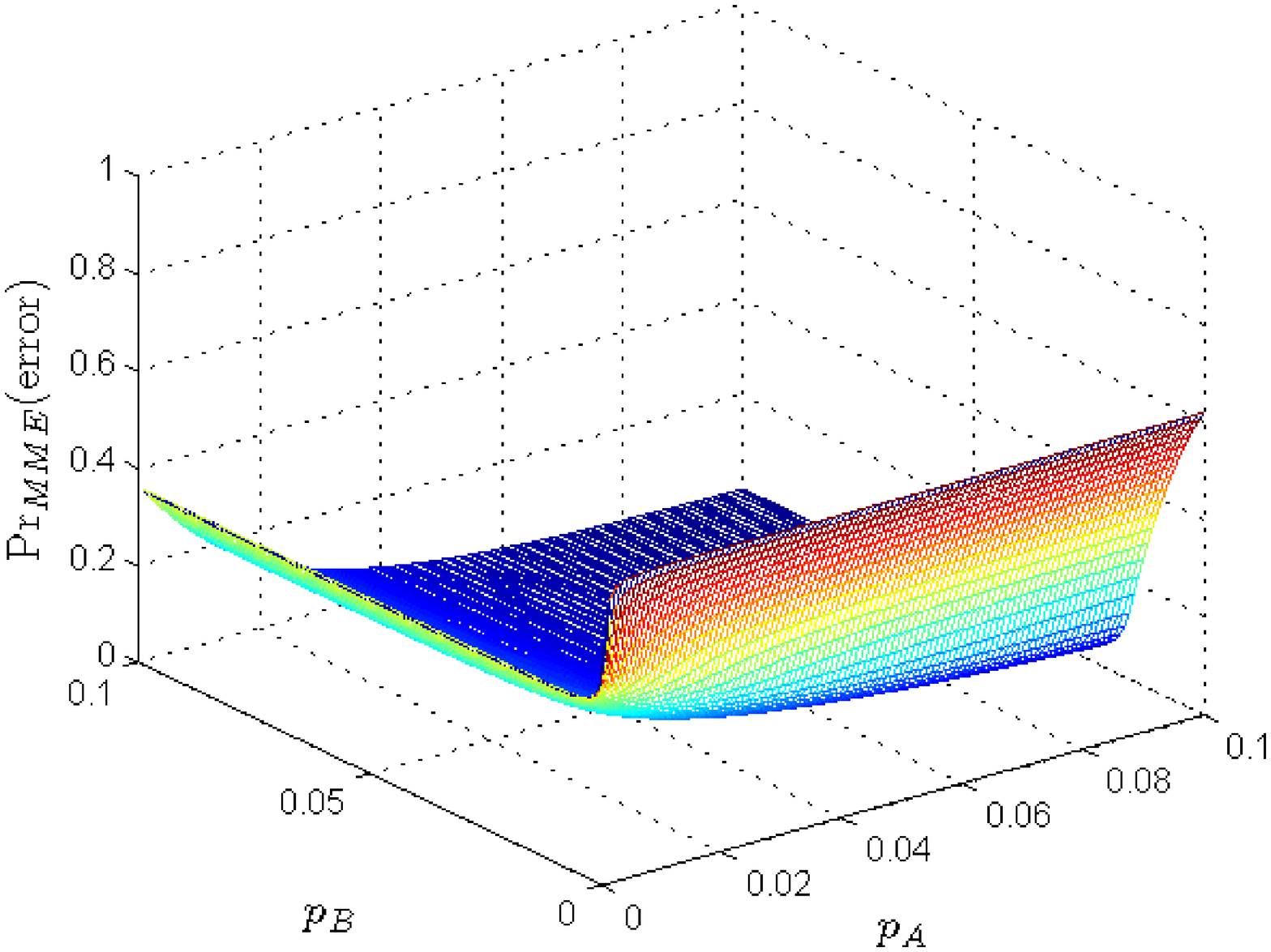}}
      }
          \caption{Analytical results of $\mbox{Pr}(\mbox{error})$ when changing $p_{\mbox{\tiny A}}$ and $p_{\mbox{\tiny B}}$ ($\tau$ = 50 time units).}
   \label{fig:error}
\end{center}
\vspace{-0.2cm}
\end{figure*}

To calculate the probability of error detection for different estimators, we first define a new random variable
\begin{equation}
Z = \delta_{\mbox{\tiny 0A}}-\delta_{\mbox{\tiny 0B}},
\end{equation}
and calculate its probability density function (pdf) $f_{\mbox{\tiny
Z}}(z)$. From Equation (\ref{eqn:exp}), we can obtain the pdf of
$\delta_{\mbox{\tiny 0B}}' = - \delta_{\mbox{\tiny 0B}}$, which is
\begin{equation}
f_{\delta_{\mbox{\tiny 0B}}'}(x) =
\left\{\begin{array} {l l} p_{\mbox{\tiny B}}\,e^{p_{\mbox{\tiny B}}x}, & \quad  x \leq 0\\
0, & \quad  x > 0.\\
\end{array} \right.
\end{equation}
Since $\delta_{\mbox{\tiny 0A}}$ and $\delta_{\mbox{\tiny 0B}}'$ are
independent, the pdf of $Z = \delta_{\mbox{\tiny
0A}}+\delta_{\mbox{\tiny 0B}}'$ is given by the convolution of
$f_{\delta_{\mbox{\tiny 0A}}}(x)$ and $f_{\delta_{\mbox{\tiny
0B}}'}(x)$, {\em i.e.,}
\begin{equation}
f_{\mbox{\tiny Z}}(z)= \int_{ - \infty }^{ + \infty }
f_{\delta_{\mbox{\tiny 0A}}}(x)f_{\delta_{\mbox{\tiny
0B}}'}(z-x)\,dx.
\end{equation}
For $z \geq 0$, this yields
\begin{eqnarray}
f_{\mbox{\tiny Z}}(z) &=& \int_{ z}^{  + \infty }
p_{\mbox{\tiny A}}\,e^{-p_{\mbox{\tiny A}}x} \cdot p_{\mbox{\tiny B}}\,e^{p_{\mbox{\tiny B}}(z-x)} \,dx \nonumber\\
&=&   \textstyle \frac{p_{\mbox{\tiny A}}p_{\mbox{\tiny B}}}{p_{\mbox{\tiny A}}+p_{\mbox{\tiny B}}}\,e^{-p_{\mbox{\tiny A}}z}.
\end{eqnarray}
For $z < 0$, we obtain
\begin{eqnarray}
f_{\mbox{\tiny Z}}(z) &=& \int_{0}^{  + \infty }
p_{\mbox{\tiny A}}\,e^{-p_{\mbox{\tiny A}}x} \cdot p_{\mbox{\tiny B}}\,e^{p_{\mbox{\tiny B}}(z-x)} \,dx \nonumber\\
&=&   \textstyle \frac{p_{\mbox{\tiny A}}p_{\mbox{\tiny B}}}{p_{\mbox{\tiny A}}+p_{\mbox{\tiny B}}}\,e^{p_{\mbox{\tiny B}}z}.
\end{eqnarray}
Hence,
\begin{equation}
\setlength{\extrarowheight}{0.15cm}
f_{\mbox{\tiny Z}}(z) =
\left\{\begin{array} {l l}
\frac{p_{\mbox{\tiny A}}p_{\mbox{\tiny B}}}{p_{\mbox{\tiny A}}+p_{\mbox{\tiny B}}}\,e^{-p_{\mbox{\tiny A}}z}, & z \geq 0\\
\frac{p_{\mbox{\tiny A}}p_{\mbox{\tiny B}}}{p_{\mbox{\tiny A}}+p_{\mbox{\tiny B}}}\,e^{p_{\mbox{\tiny B}}z}, &  z< 0.\\
\end{array} \right.
\end{equation}

\begin{figure*}[tb]
\begin{center}
    \mbox{
       \subfigure[$\mbox{Pr}(\mbox{error})$ ($p_{\mbox{\tiny A}}=0.02$ and $p_{\mbox{\tiny B}}=0.05$).]{\includegraphics[width=2.5in]{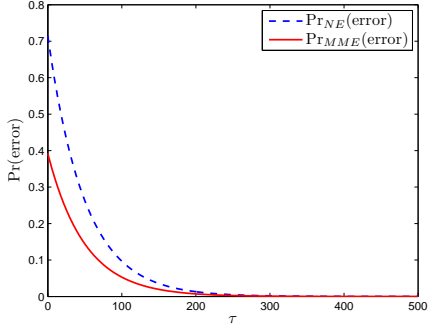}}
  \hspace{3cm}
         \subfigure[$\mbox{Pr}(\mbox{error})$ ($p_{\mbox{\tiny A}}=0.05$ and $p_{\mbox{\tiny B}}=0.02$).]{\includegraphics[width=2.5in]{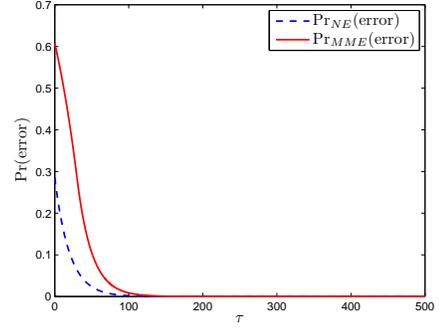}}
      }
         \caption{Analytical results of $\mbox{Pr}(\mbox{error})$ when changing $\tau$.}
   \label{fig:error2}
\end{center}
\vspace{-0.3cm}
\end{figure*}

\subsubsection{Naive Estimator}

The naive estimator uses $\hat{t_{\mbox{\tiny 0}}}= t_{\mbox {\tiny
1}}-1$ to estimate $t_{\mbox {\tiny 0}}$. Thus, the probability of
error detection is
\begin{equation}
\mbox{Pr}_{\mbox {\tiny NE}}(\mbox{error})
 = \mbox{Pr}(t_{\mbox{\tiny 1A}}-1 > t_{\mbox {\tiny 1B}}-1)
 = \mbox{Pr}(\delta_{\mbox{\tiny 0A}} > \tau+\delta_{\mbox{\tiny 0B}}),
\end{equation}
where $\tau=t_{\mbox{\tiny 0B}}-t_{\mbox{\tiny 0A}}$, the time interval between the infection of host A and host B; and $\tau>0$.
We then have
\begin{eqnarray}
\mbox{Pr}_{\mbox {\tiny NE}}(\mbox{error})
& = & \mbox{Pr}(\delta_{\mbox{\tiny 0A}}-\delta_{\mbox{\tiny 0B}} >\tau)  \nonumber \\
& = & \mbox{Pr}(Z > \tau) \nonumber \\
& = &\int_{\tau }^{+\infty} \textstyle \frac{p_{\mbox{\tiny A}}p_{\mbox{\tiny B}}}{p_{\mbox{\tiny A}}+p_{\mbox{\tiny B}}}\,e^{-p_{\mbox{\tiny A}}z}\, dz  \nonumber \\
\label{equ:error_NE}
&=&\textstyle \frac{p_{\mbox{\tiny
B}}}{p_{\mbox{\tiny A}}+p_{\mbox{\tiny B}}}\,e^{-p_{\mbox{\tiny
A}}\tau}.
\end{eqnarray}
Note that another way to derive $\mbox{Pr}_{\mbox {\tiny
NE}}(\mbox{error})$ is based on the memoryless property of the
exponential distribution and $\mbox{Pr}(\delta_{\mbox{\tiny 0A}} >
\delta_{\mbox{\tiny 0B}})=p_{\mbox{\tiny B}}/(p_{\mbox{\tiny
A}}+p_{\mbox{\tiny B}})$, {\em i.e.,}
\begin{equation}
\mbox{Pr}_{\mbox {\tiny NE}}(\mbox{error}) =
\mbox{Pr}(\delta_{\mbox{\tiny 0A}} > \tau+\delta_{\mbox{\tiny 0B}})
= \mbox{Pr}(\delta_{\mbox{\tiny 0A}} > \tau)
\mbox{Pr}(\delta_{\mbox{\tiny 0A}} > \delta_{\mbox{\tiny 0B}}),
\end{equation}
which leads to the same result.

\subsubsection{Proposed Estimators}
We assume that Darknet observes a sufficient number of scans from
hosts A and B so that our proposed estimators can estimate
$\mu_{\mbox{\tiny A}}$ ({\em i.e.,} $\frac{1}{p_{\mbox{\tiny A}}}$) and
$\mu_{\mbox{\tiny B}}$ ({\em i.e.,} $\frac{1}{p_{\mbox{\tiny B}}}$)
accurately. Then, the probability of error detection of our proposed
estimators is
\begin{eqnarray}
\mbox{Pr}_{\mbox {\tiny MME}}(\mbox{error})&=&\mbox{Pr}_{\mbox {\tiny MLE}}(\mbox{error}) = \mbox{Pr}_{\mbox {\tiny LRE}}(\mbox{error}) \nonumber \\
&=& \mbox{Pr}(t_{\mbox {\tiny 1A}}- \textstyle {1\over p_{\mbox {\tiny A}}} > t_{\mbox{\tiny 1B}}-{1\over p_{\mbox {\tiny B}}}) \nonumber \\
&=& \mbox{Pr}(\delta_{\mbox {\tiny 0A}}-\delta_{\mbox{\tiny 0B}}>\tau + \textstyle {1\over p_{\mbox {\tiny A}}}-\textstyle {1\over p_{\mbox {\tiny B}}}) \nonumber \\
&=&\textstyle\mbox{Pr}(Z > \tau + \frac{p_{\mbox{\tiny B}}-p_{\mbox{\tiny A}}}{p_{\mbox{\tiny A}}p_{\mbox{\tiny B}}})  \nonumber \\
&=& \int_{\tau+\frac{p_{\mbox{\tiny B}}-p_{\mbox{\tiny
A}}}{p_{\mbox{\tiny A}}p_{\mbox{\tiny B}}} }^{+\infty}
f_{\mbox{\tiny Z}}(z) \, dz.
\end{eqnarray}
When $\tau+\frac{p_{\mbox{\tiny B}}-p_{\mbox{\tiny A}}}{p_{\mbox{\tiny A}}p_{\mbox{\tiny B}}} \geq 0$,
\begin{eqnarray}
\mbox{Pr}_{\mbox {\tiny MME}}(\mbox{error}) & =
&\int_{\tau+\frac{p_{\mbox{\tiny B}}-p_{\mbox{\tiny
A}}}{p_{\mbox{\tiny A}}p_{\mbox{\tiny B}}} }^{+\infty}
\textstyle \frac{p_{\mbox{\tiny A}}p_{\mbox{\tiny B}}}{p_{\mbox{\tiny A}}+p_{\mbox{\tiny B}}}\,e^{-p_{\mbox{\tiny A}}z}\, dz \nonumber \\
&=&\textstyle \frac{p_{\mbox{\tiny B}}}{p_{\mbox{\tiny
A}}+p_{\mbox{\tiny B}}}\,e^{-p_{\mbox{\tiny
A}}\big(\tau+\frac{p_{\mbox{\tiny B}}-p_{\mbox{\tiny
A}}}{p_{\mbox{\tiny A}}p_{\mbox{\tiny B}}}\big)}. \label{eqn:pmme1}
\end{eqnarray}
When $\tau+\frac{p_{\mbox{\tiny B}}-p_{\mbox{\tiny A}}}{p_{\mbox{\tiny A}}p_{\mbox{\tiny B}}} < 0$,
\begin{eqnarray}
\mbox{Pr}_{\mbox {\tiny MME}}(\mbox{error}) & =
&\int_{\tau+\frac{p_{\mbox{\tiny B}}-p_{\mbox{\tiny
A}}}{p_{\mbox{\tiny A}}p_{\mbox{\tiny B}}} }^{0}
\textstyle \frac{p_{\mbox{\tiny A}}p_{\mbox{\tiny B}}}{p_{\mbox{\tiny A}}+p_{\mbox{\tiny B}}}\,e^{p_{\mbox{\tiny B}}z}\, dz + \nonumber \\
& & \int_{0 }^{+\infty} \textstyle \frac{p_{\mbox{\tiny A}}p_{\mbox{\tiny B}}}{p_{\mbox{\tiny A}}+p_{\mbox{\tiny B}}}\,e^{-p_{\mbox{\tiny A}}z}\, dz  \nonumber \\
& = & \!\!\!\! \textstyle \frac{1}{p_{\mbox{\tiny A}}+p_{\mbox{\tiny
B}}}\Big(p_{\mbox{\tiny A}}+p_{\mbox{\tiny B}}-p_{\mbox{\tiny
A}}\,e^{p_{\mbox{\tiny B}}\big(\tau+\frac{p_{\mbox{\tiny
B}}-p_{\mbox{\tiny A}}}{p_{\mbox{\tiny A}}p_{\mbox{\tiny
B}}}\big)}\Big). \label{eqn:pmme2}
\end{eqnarray}

\subsubsection{Performance Comparison}

Since $\mbox{Pr}_{\mbox {\tiny NE}}(\mbox{error}) =
\mbox{Pr}(Z>\tau)$ and $\mbox{Pr}_{\mbox {\tiny MME}}(\mbox{error})
= \textstyle\mbox{Pr}\big(Z > \tau+\frac{p_{\mbox{\tiny
B}}-p_{\mbox{\tiny A}}}{p_{\mbox{\tiny A}}p_{\mbox{\tiny B}}}\big)$,
for a given $\tau$ ($\tau > 0$), comparing Equation
(\ref{equ:error_NE}) with Equations (\ref{eqn:pmme1}) and
(\ref{eqn:pmme2}),
\begin{equation}
\left\{\begin{array} {l l}
\mbox{Pr}_{\mbox {\tiny NE}}(\mbox{error}) > \mbox{Pr}_{\mbox {\tiny MME}}(\mbox{error}), &  p_{\mbox{\tiny A}} < p_{\mbox{\tiny B}} \\
\mbox{Pr}_{\mbox {\tiny NE}}(\mbox{error}) < \mbox{Pr}_{\mbox {\tiny MME}}(\mbox{error}), &  p_{\mbox{\tiny A}} > p_{\mbox{\tiny B}}.\\
\end{array} \right.
\end{equation}
Hence, it is unclear which estimator is better based on the
expressions of $\mbox{Pr}_{\mbox {\tiny NE}}(\mbox{error})$ and
$\mbox{Pr}_{\mbox {\tiny MME}}(\mbox{error})$.
However, we can compare the performance of our estimators with the
naive estimator through numerical analysis. We first demonstrate the probabilities of error
detection ({\em i.e.,} $\mbox{Pr}_{\mbox {\tiny NE}}(\mbox{error})$
and $\mbox{Pr}_{\mbox {\tiny MME}}(\mbox{error})$) as the functions
of $p_{\mbox{\tiny A}}$ and $p_{\mbox{\tiny B}}$ in Fig.
\ref{fig:error}, where $\tau$ = 50 time units. It can be seen that
for the naive estimator, when host A hits the Darknet with a very
low probability, $\mbox{Pr}_{\mbox {\tiny NE}}(\mbox{error})$ is
almost 1 regardless of $p_{\mbox{\tiny B}}$. However, the worst case
of $\mbox{Pr}_{\mbox {\tiny MME}}(\mbox{error})$ is slightly above
0.6 when $p_{\mbox{\tiny B}}$ is small. Moreover, we show the
probabilities of error detection as a function of $\tau$ with a
given pair of $ p_{\mbox{\tiny A}}$ and $ p_{\mbox{\tiny B}}$ in
Fig. \ref{fig:error2}. The performance of two estimators improves as
$\tau$ increases. Furthermore, the sum of the integral
$\int_{0}^{500 } \mbox{Pr}_{\mbox {\tiny NE}}(\mbox{error}) \, d
\tau $ of the two figures is 41.43, while the sum of the integral
$\int_{0}^{500 } \mbox{Pr}_{\mbox {\tiny MME}}(\mbox{error}) \, d
\tau $ in these two cases is only 34.76. This shows that the
improvement gain of our estimators over the naive estimator when
$p_{\mbox{\tiny A}} < p_{\mbox{\tiny B}}$ outweighs the
degradation suffered when $p_{\mbox{\tiny A}} > p_{\mbox{\tiny B}}$,
indicating the benefits of applying our estimators.

\begin{figure*}[htb]
\begin{center}
    \mbox{
\hspace{-0.8cm}
      \subfigure[Comparison of $\hat{\mu}$.]{\includegraphics[width=2.25in]{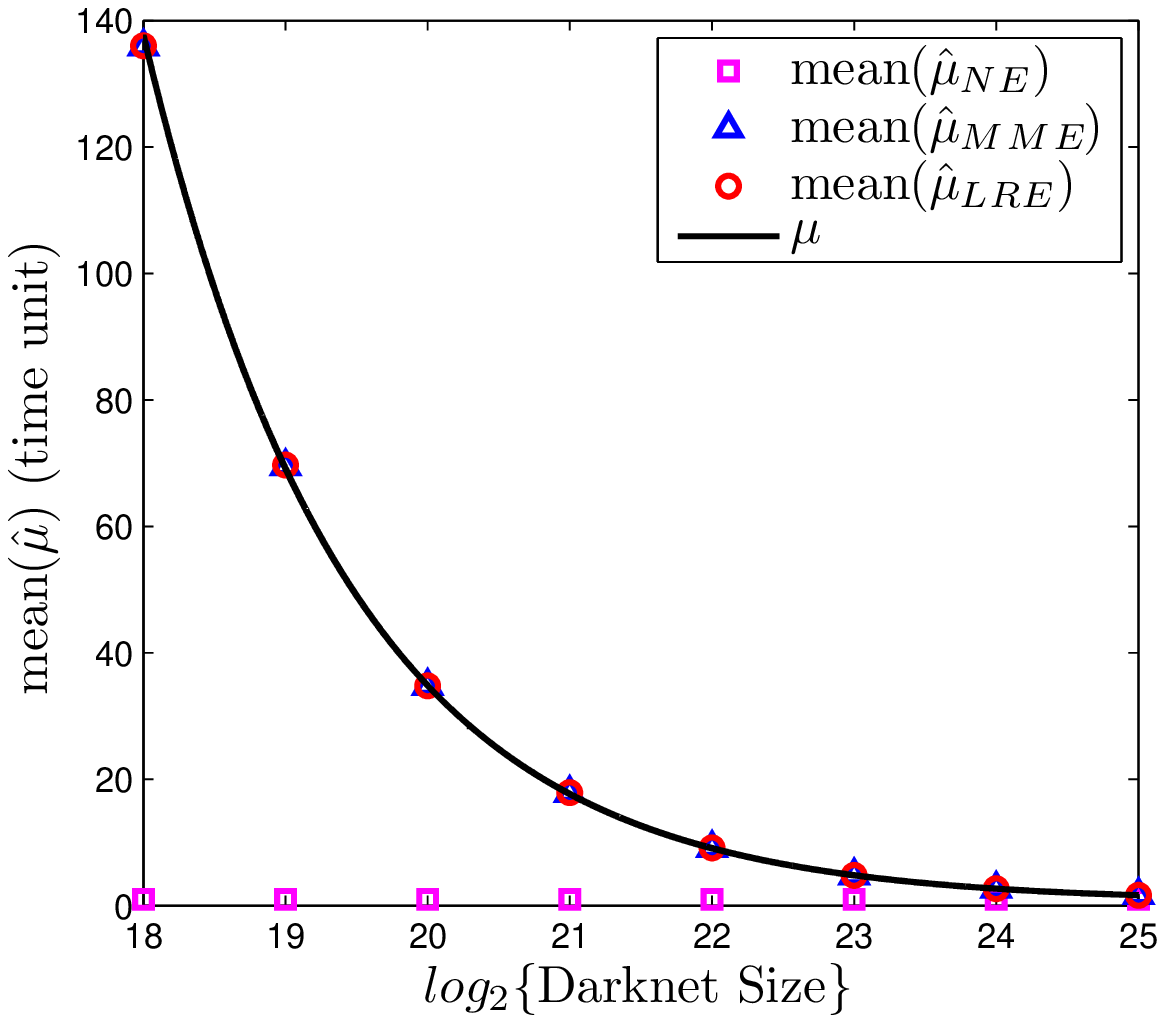}}
      \subfigure[Comparison of $\hat{t_0}$.]{\includegraphics[width=2.25in]{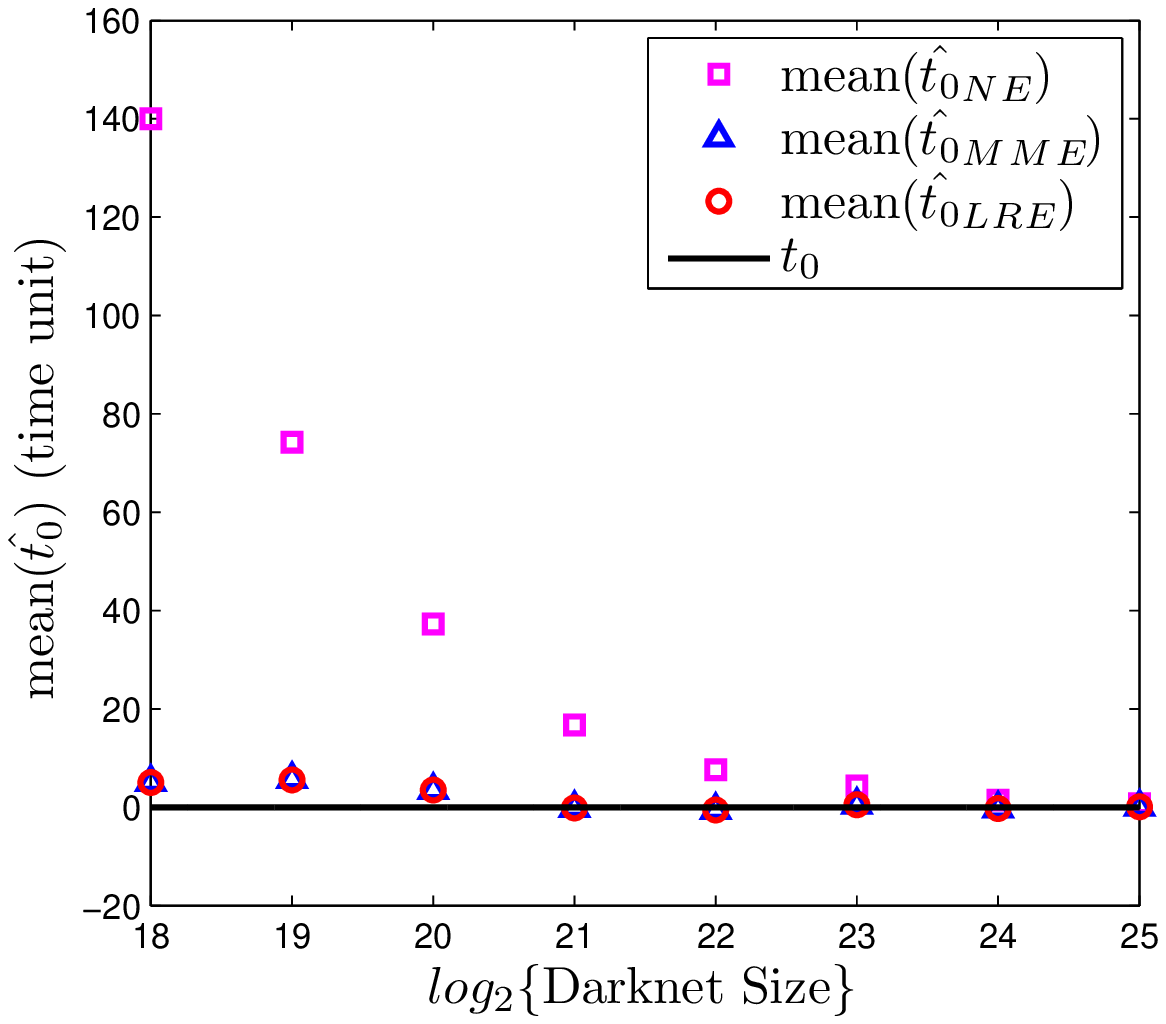}}
      \subfigure[Comparison of $\mbox{MSE}(\hat{t_0})$.]{\includegraphics[width=2.25in]{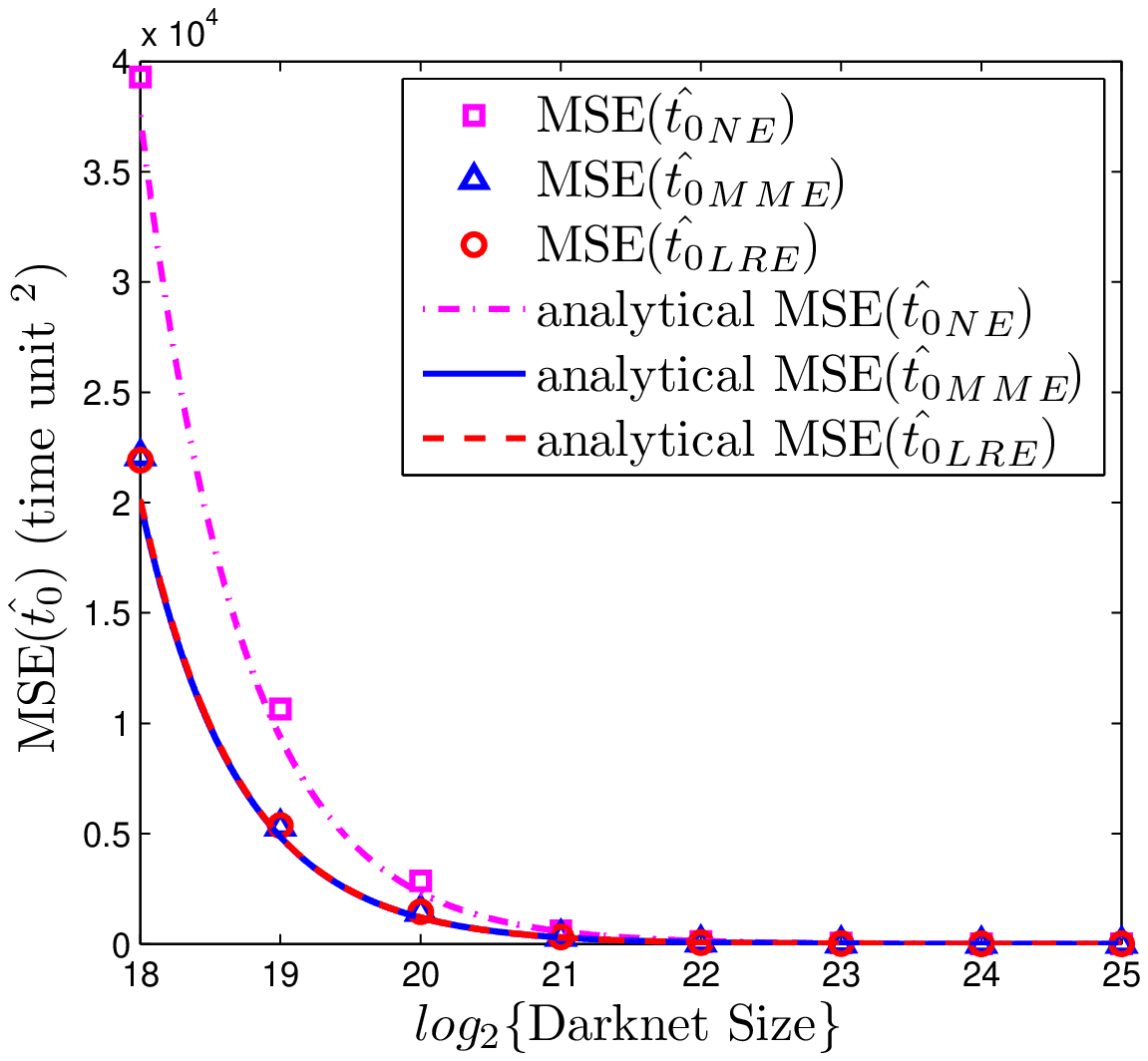}}
      }
    \caption{Simulation results of changing the Darknet size for random scanning (all cases are for scanning rate: 358 scans/min, observation window size: 800 mins).}
\label{size}
\end{center}
\end{figure*}
\begin{figure*}[htb]
\begin{center}
    \mbox{
\hspace{-0.8cm}
      \subfigure[Comparison of $\hat{\mu}$.]{\includegraphics[width=2.25in]{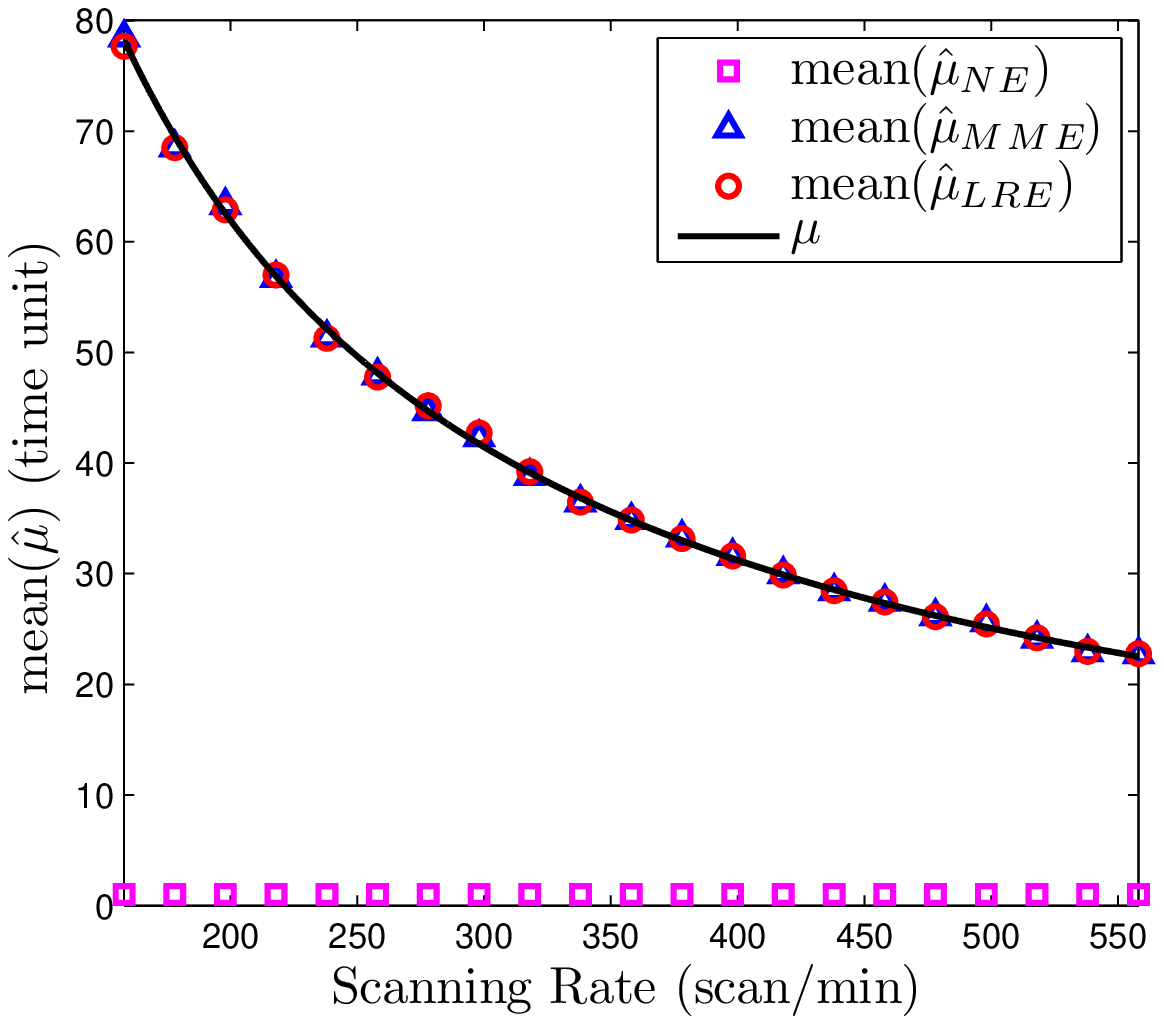}}
      \subfigure[Comparison of $\hat{t_0}$.]{\includegraphics[width=2.25in]{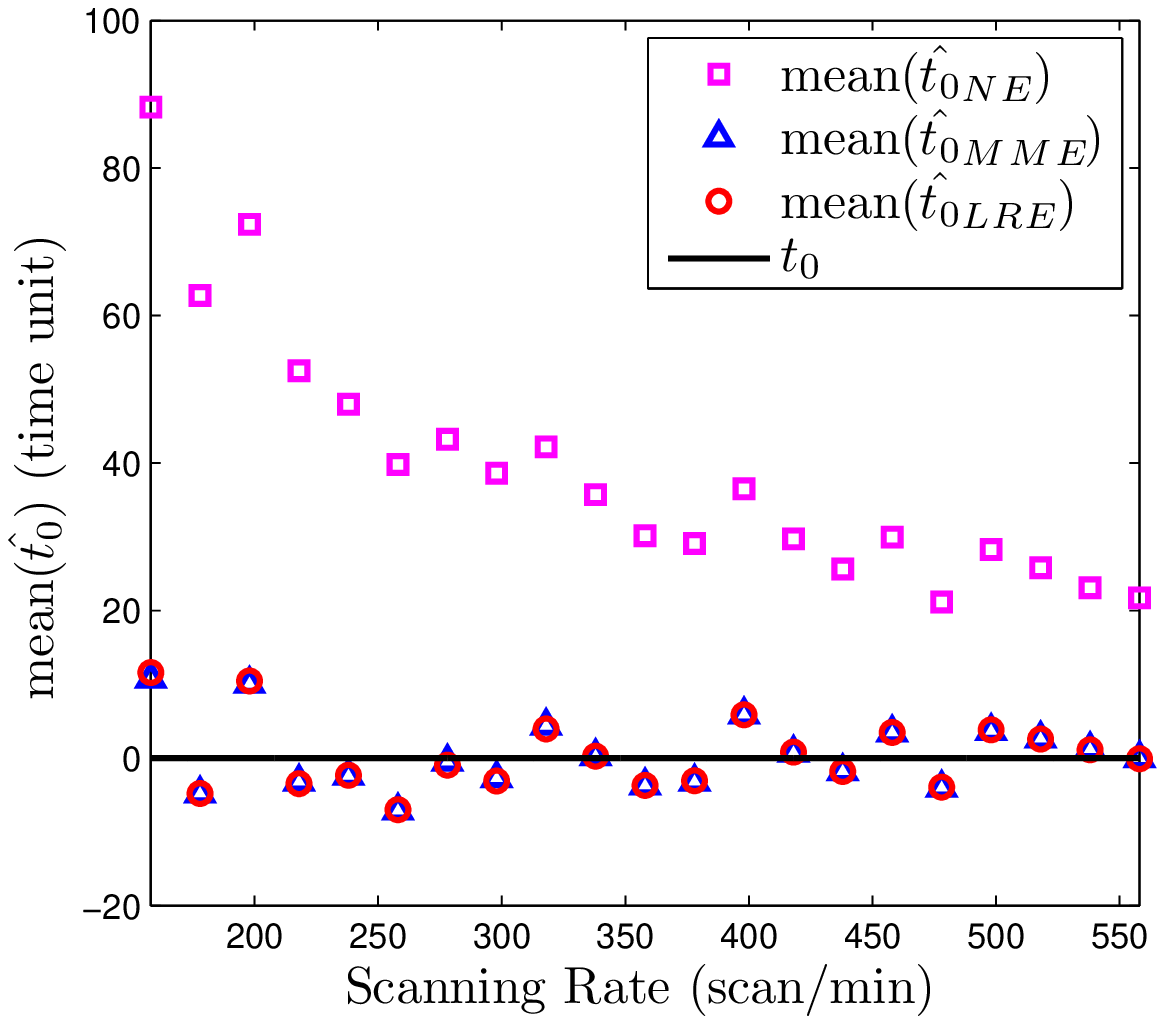}}
      \subfigure[Comparison of $\mbox{MSE}(\hat{t_0})$.]{\includegraphics[width=2.25in]{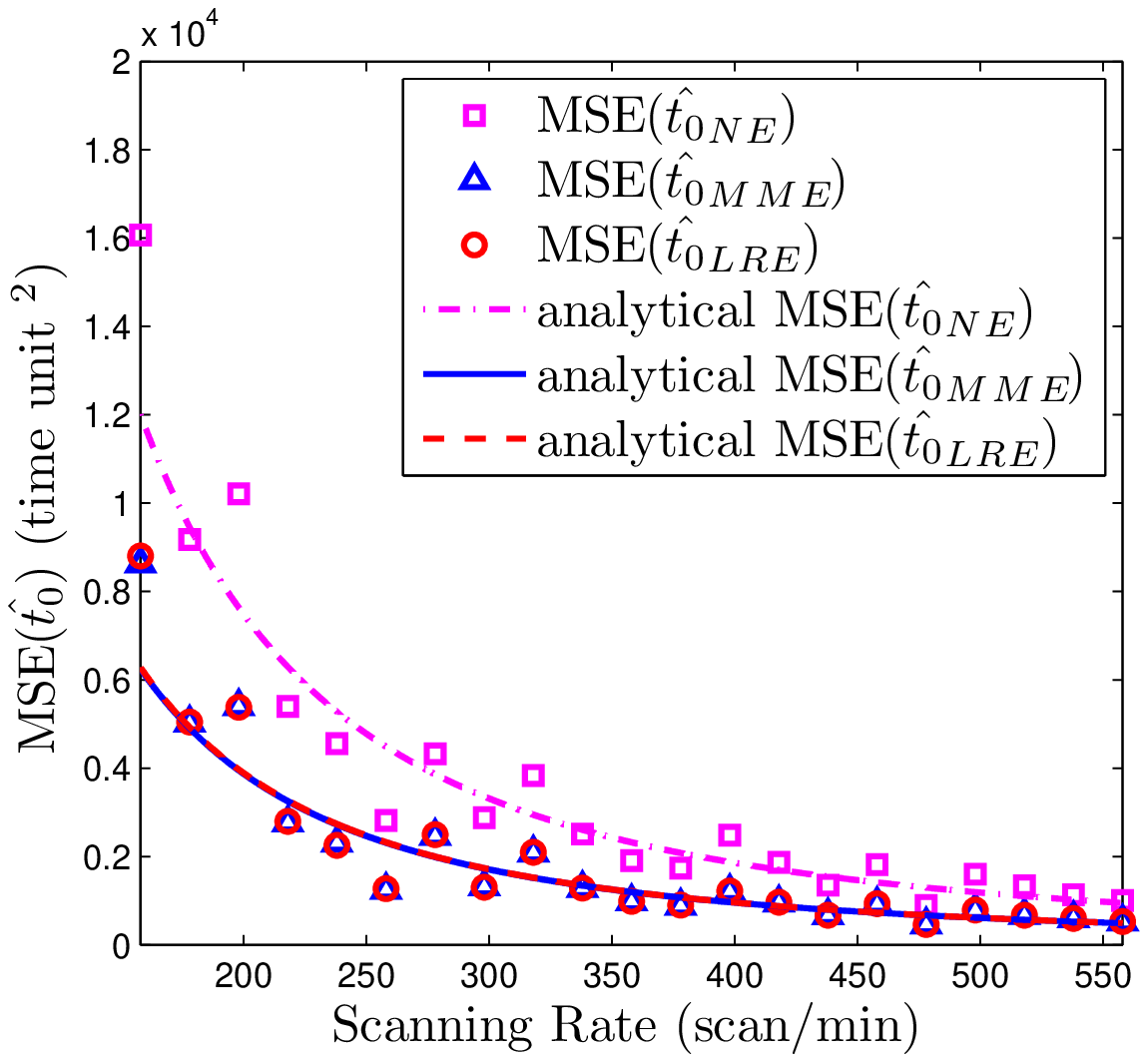}}
      }
    \caption{Simulation results of changing the scanning rate for random scanning (all cases are for Darknet size: $2^{20}$ IP addresses, observation window size: 800 mins).}
   \label{rate}
   \end{center}
\end{figure*}
\begin{figure*}[htb]
\begin{center}
    \mbox{
    \hspace{-0.8cm}
      \subfigure[Comparison of $\hat{\mu}$.]{\includegraphics[width=2.25in]{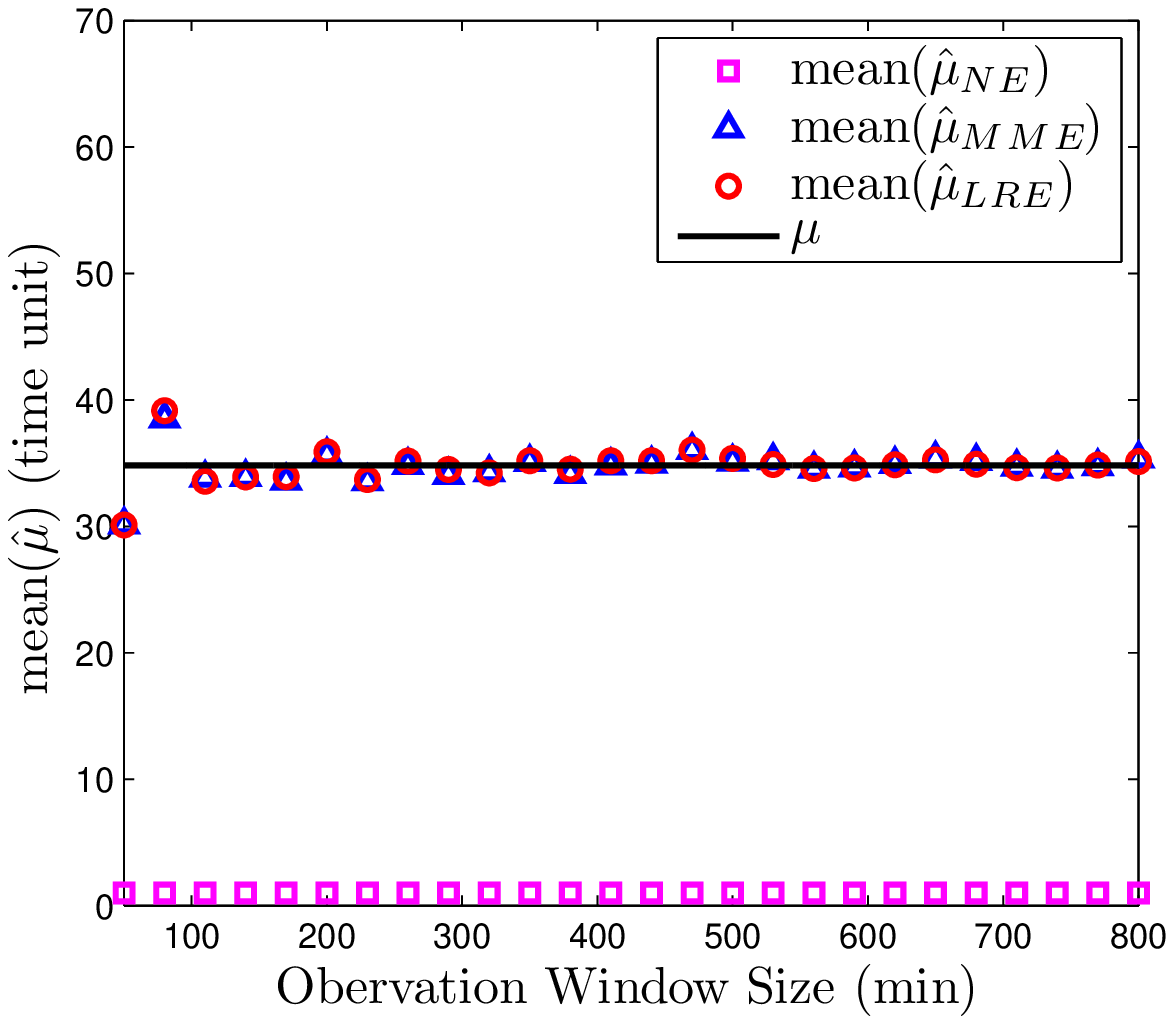}}
      \subfigure[Comparison of $\hat{t_0}$.]{\includegraphics[width=2.25in]{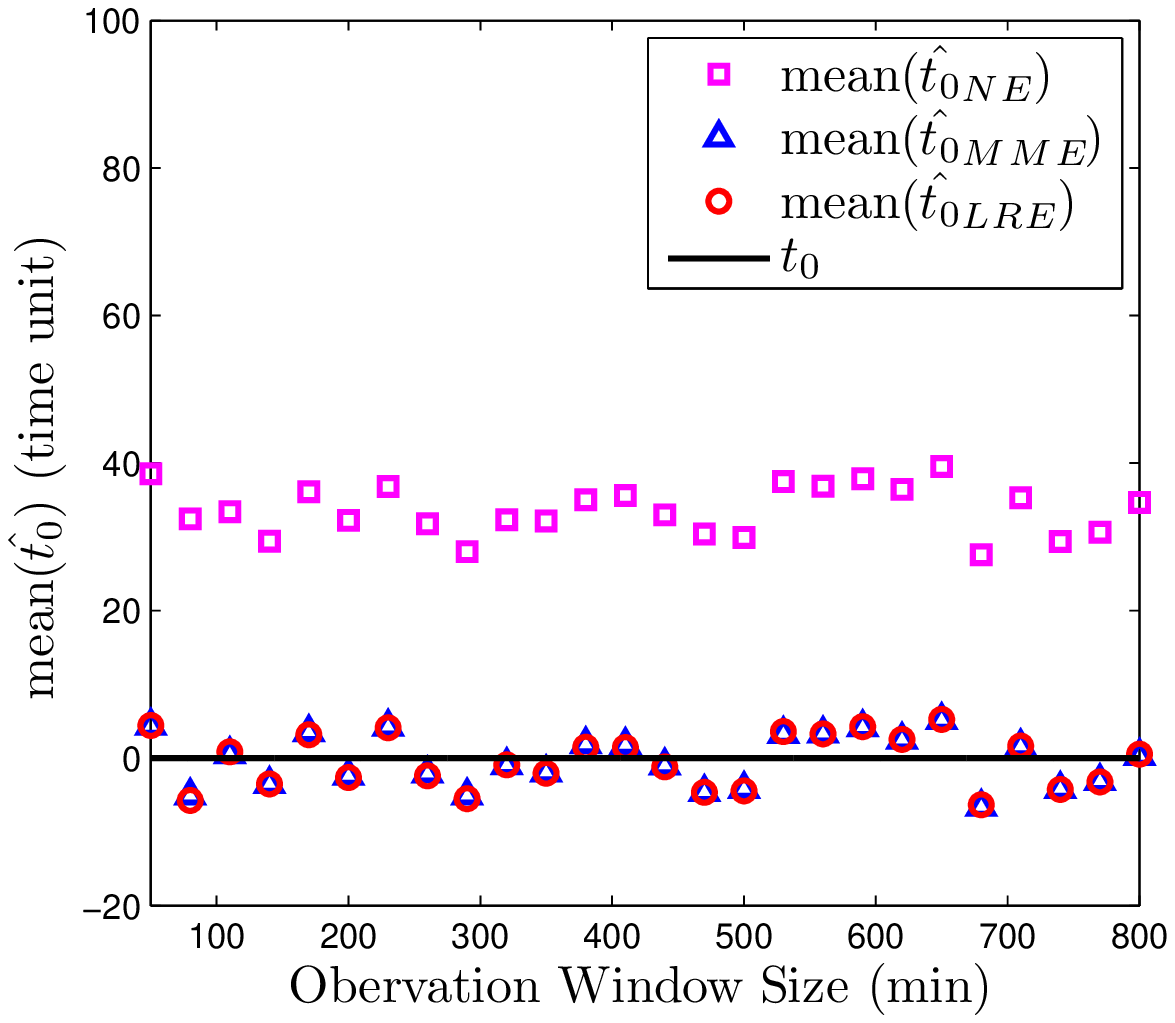}}
      \subfigure[Comparison of $\mbox{MSE}(\hat{t_0})$.]{\includegraphics[width=2.25in]{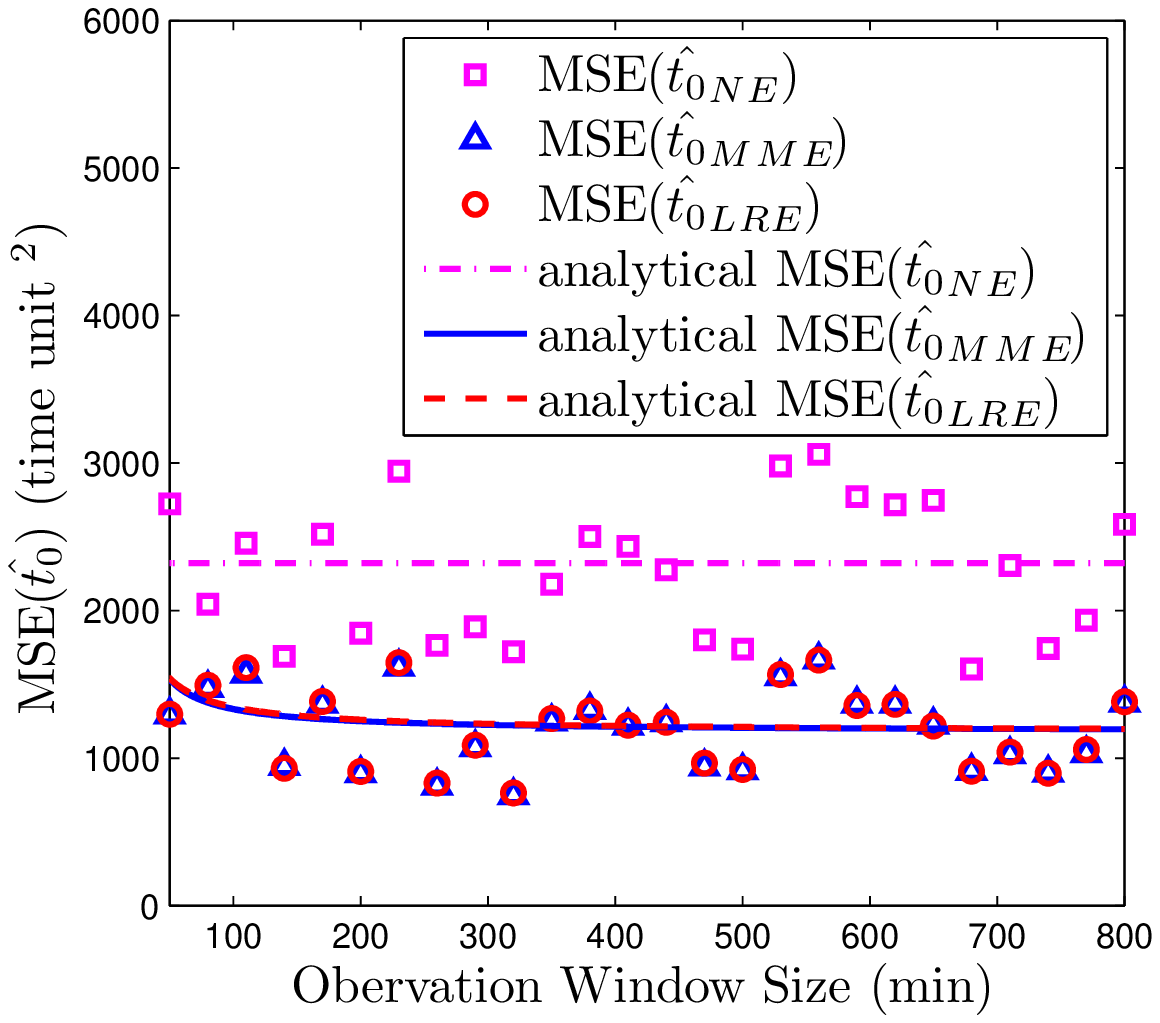}}
      }
    \caption{Simulation results of changing the observation window size for random scanning (all cases are for scanning rate: 358 scans/min, Darknet size: $2^{20}$ IP addresses).}
\label{win}
\end{center}
\vspace{-0.3cm}
\end{figure*}
\begin{figure*}[htb]
\begin{center}
    \mbox{
    \hspace{-0.8cm}
      \subfigure[Comparison of $\mbox{MSE}(\hat{t_0})$.]{\includegraphics[width=2.25in]{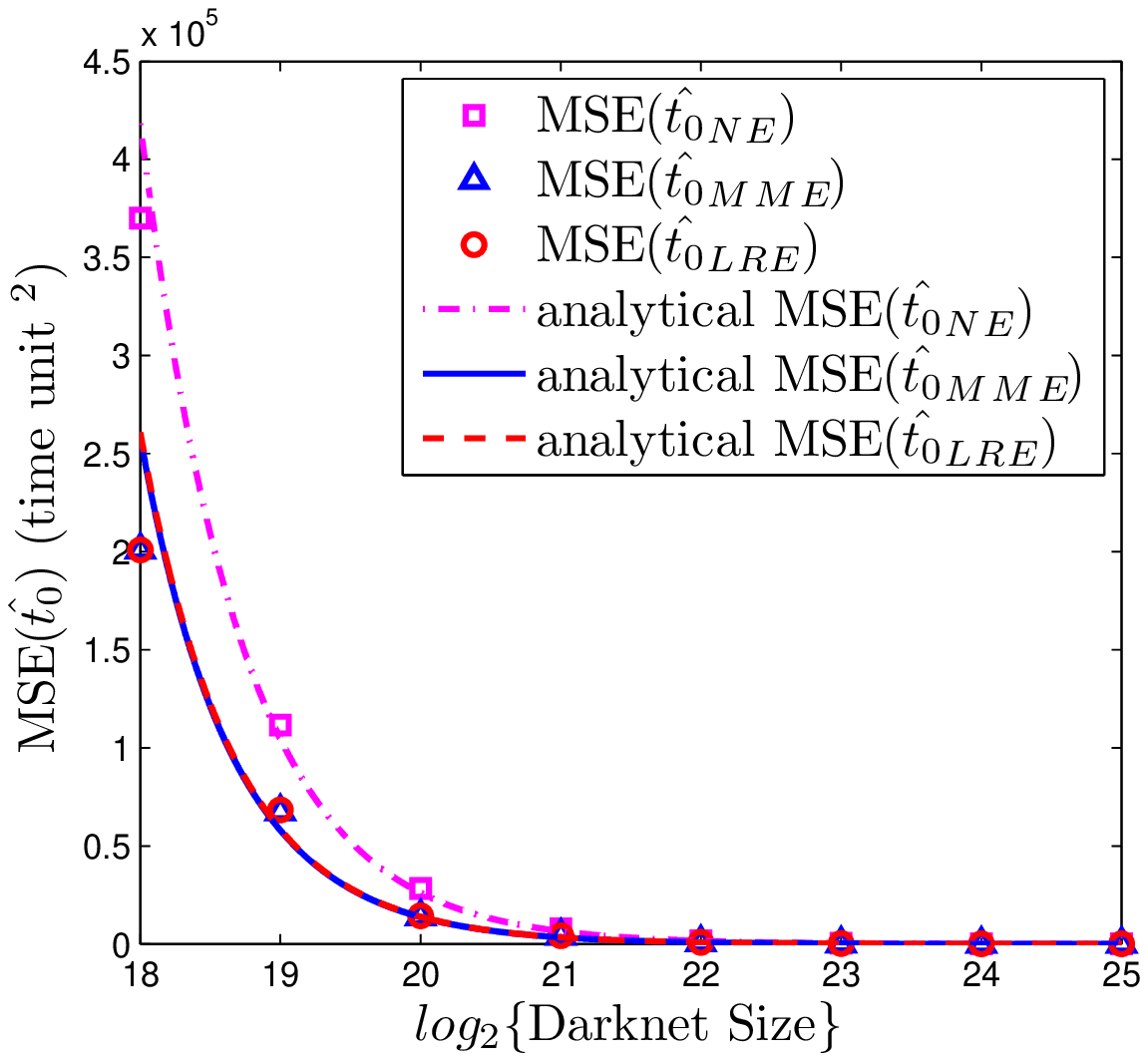}}
      \subfigure[Comparison of $\mbox{MSE}(\hat{t_0})$.]{\includegraphics[width=2.25in]{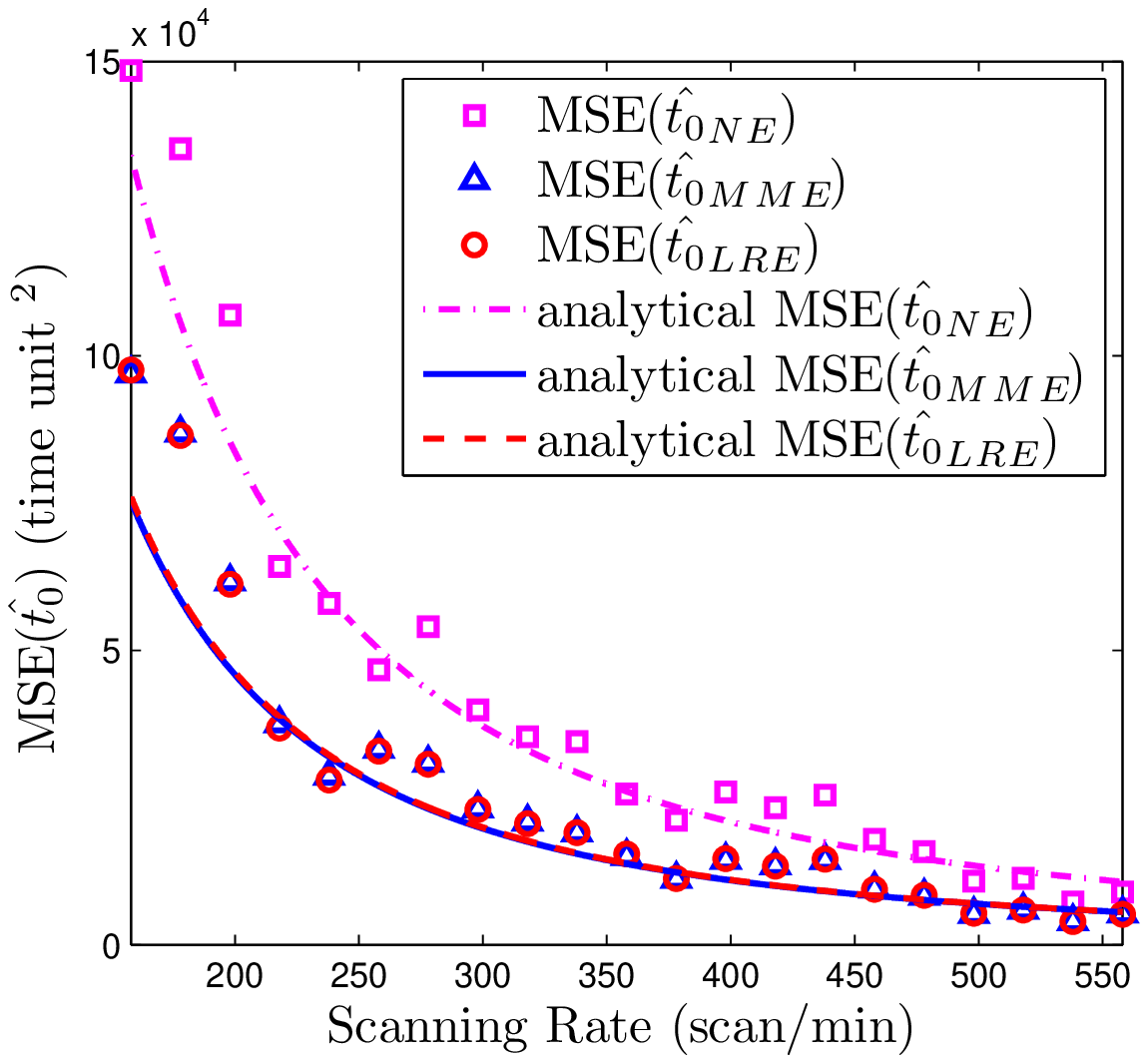}}
      \subfigure[Comparison of $\mbox{MSE}(\hat{t_0})$.]{\includegraphics[width=2.25in]{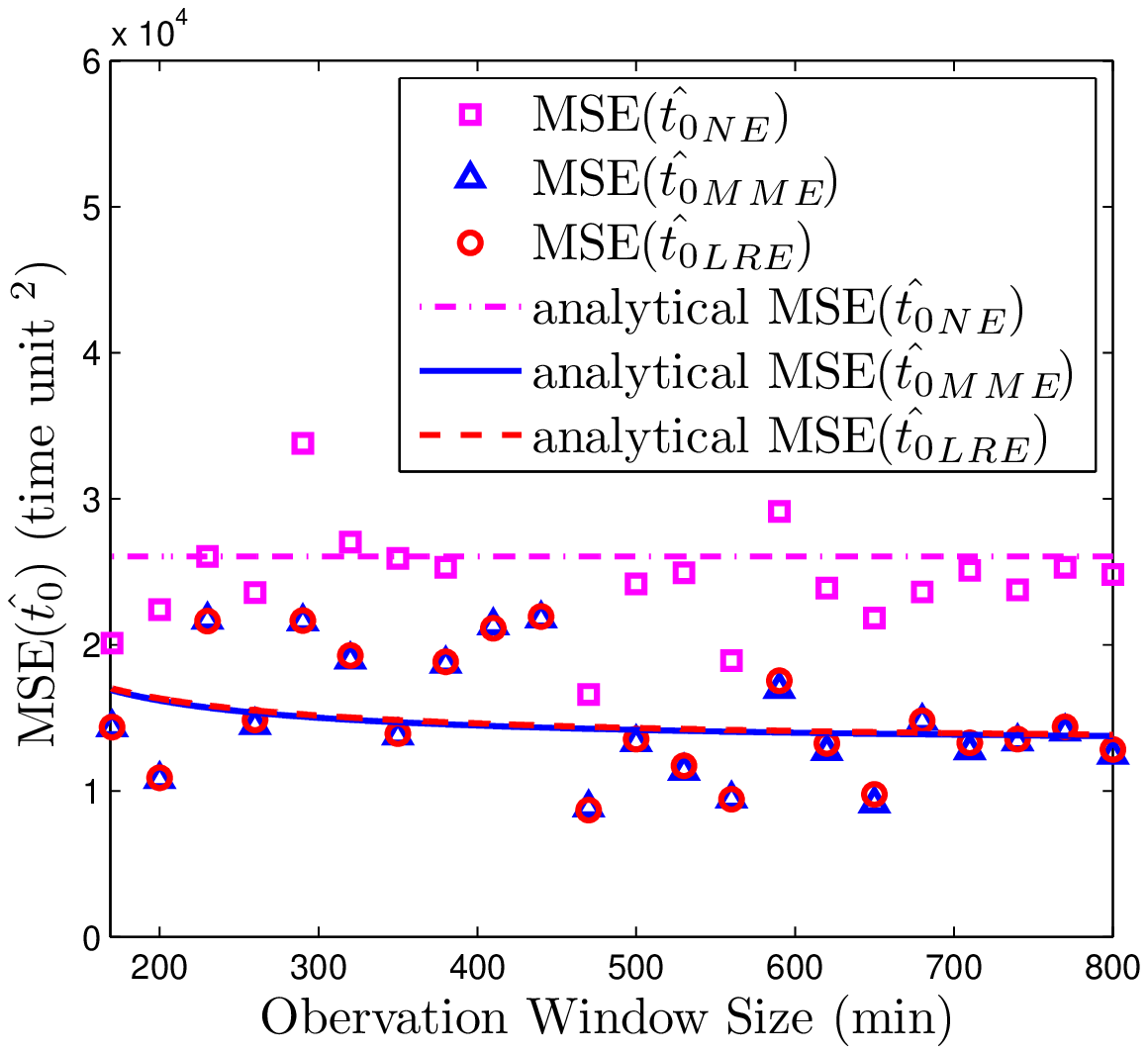}}
      }
    \caption{(a) Simulation results of changing the Darknet size for localized scanning ($p_a$ = 0.7, scanning rate: 358 scans/min, observation window size: 800 mins).
    (b) Simulation results of changing the scanning rate for localized scanning ($p_a$ = 0.7, Darknet size: $2^{20}$ IP addresses, observation window size: 800 mins).
    (c) Simulation results of changing the observation window size for localized scanning ($p_a$ = 0.7, scanning rate: 358 scans/min, Darknet size: $2^{20}$ IP addresses).}
\label{fig:Lmse}
\end{center}
\vspace{-0.3cm}
\end{figure*}

Note that $p_{\mbox{\tiny A}}$, $p_{\mbox{\tiny B}}$, and $\tau$ can
be random variables. To evaluate the overall performance of each
estimator, we consider the average probability of error
detection over $p_{\mbox{\tiny A}}$, $p_{\mbox{\tiny B}}$, and
$\tau$, {\em i.e.,}
\begin{equation}
  \mbox{E}\,[\mbox{Pr}(\mbox{error})] = \int_{\tau } \int_{p_{\mbox{\tiny A}} }\int_{p_{\mbox{\tiny
B}}} \mbox{Pr}(\mbox{error}) \cdot f(p_{\mbox{\tiny
A}},p_{\mbox{\tiny B}}, \tau)
 \,\, d \, p_{\mbox{\tiny B}}\, d\, p_{\mbox{\tiny A}} \, d \tau.
\end{equation}
Since $p_{\mbox{\tiny A}}$, $p_{\mbox{\tiny B}}$, and $\tau$ are
independent,
\begin{equation}
f(p_{\mbox{\tiny A}},p_{\mbox{\tiny B}}, \tau) = f(p_{\mbox{\tiny
A}})\cdot f(p_{\mbox{\tiny B}})\cdot f(\tau).
\end{equation}
We then consider some cases in which we are interested and apply the
numerical integration toolbox in Matlab \cite{NIT} to calculate the
triple integration. For example, we assume that $s_{\mbox{\tiny A}}$
and $s_{\mbox{\tiny B}}$ follow a normal distribution $N(u,
\sigma^2)$ and $\tau$ is uniform over $(0, \tau_1]$. We find that
when $u$, $\sigma^2$, and $\tau_1$ are set to realistic values, we
always have
\begin{equation}
    \mbox{E}\,[\mbox{Pr}_{\mbox {\tiny NE}}(\mbox{error})] > \mbox{E}\,[\mbox{Pr}_{\mbox {\tiny
    MME}}(\mbox{error})].
\end{equation}
That is, our proposed estimators perform better than NE on average, which will further be verified in Section 4 through simulations.

Moreover, in Fig. 5(a), it can
be seen that the majority of detection error for the
naive estimator comes from the case that $p_{\mbox{\tiny A}} <
p_{\mbox{\tiny B}}$. Specifically, it is obvious to derive the following
theorem from Equations (34) and (37).
\begin{theorem}
\label{thm:sequence}
When $p_{\mbox{\tiny A}} < p_{\mbox{\tiny B}}$,
\begin{eqnarray}
\mbox{Pr}_{\mbox {\tiny MME}}(\mbox{error}) &= &\mbox{Pr}_{\mbox
{\tiny MLE}}(\mbox{error}) = \mbox{Pr}_{\mbox {\tiny
LRE}}(\mbox{error}) \nonumber\\
&= &\mbox{Pr}_{\mbox {\tiny NE}}(\mbox{error})\cdot e^{-\big(
1-\frac{p_{\mbox{\tiny A}}}{p_{\mbox{\tiny B}}} \big)}.
\end{eqnarray}
\end{theorem}
That is, the error probability is decreased by a factor of
$e^{-\big( 1-\frac{p_{\mbox{\tiny A}}}{p_{\mbox{\tiny B}}} \big)}$
by applying our estimators as compared with the naive estimator.

\begin{figure*}[htb]
\begin{center}
    \mbox{
    \hspace{-0.8cm}
      \subfigure[Comparison of $\hat{\mu}$.]{\includegraphics[width=2.25in]{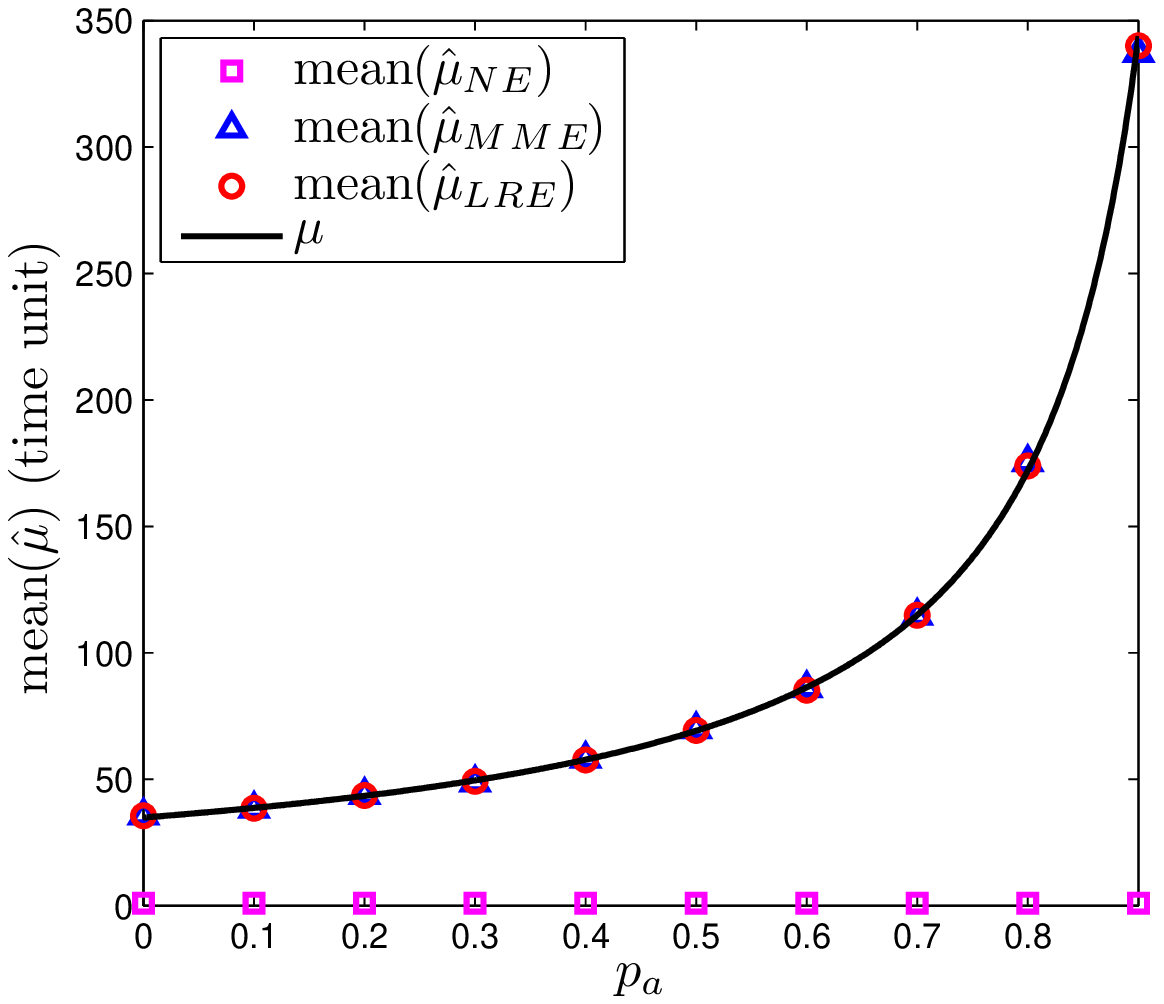}}
      \subfigure[Comparison of $\hat{t_0}$.]{\includegraphics[width=2.25in]{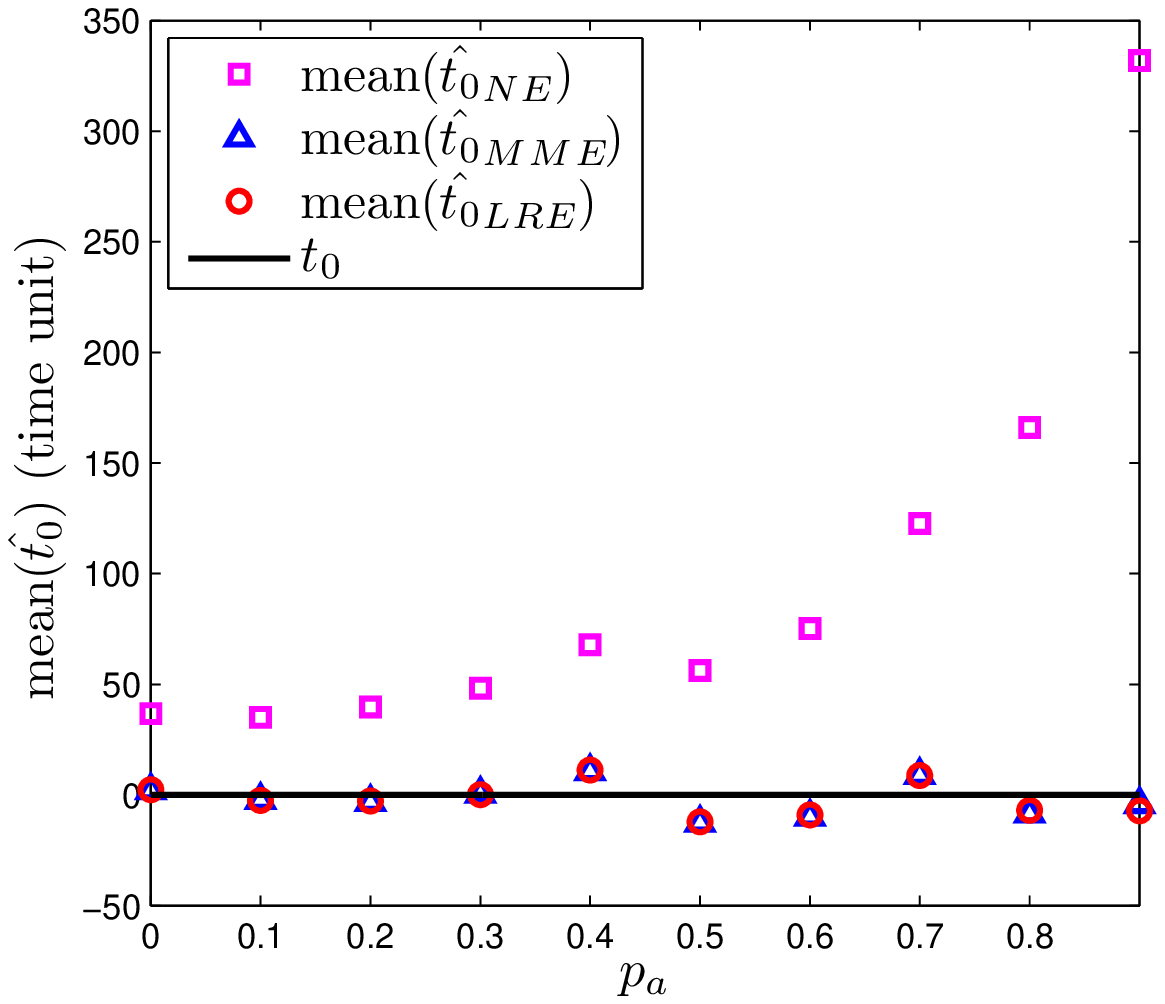}}
      \subfigure[Comparison of $\mbox{MSE}(\hat{t_0})$.]{\includegraphics[width=2.25in]{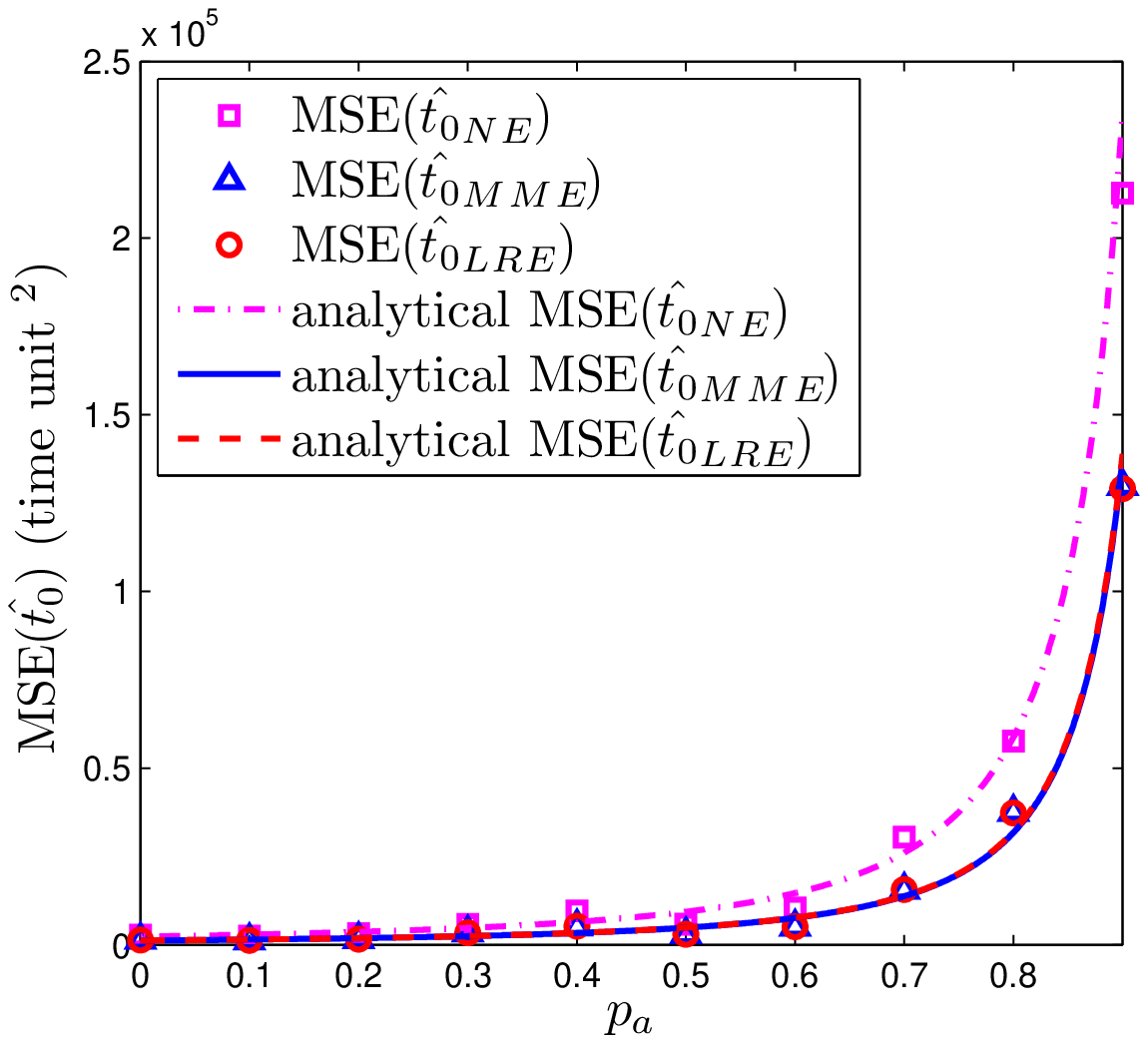}}
      }
    \caption{Simulation results of changing $p_a$ for localized scanning (all cases are for scanning rate: 358 scans/min, Darknet size: $2^{20}$ IP addresses, observation window size: 800 mins).}
\label{Lpa}
\end{center}
\end{figure*}
\begin{figure*}[htb]
\begin{center}
    \mbox{
    \hspace{-0.8cm}
      \subfigure[]{\includegraphics[width=2.25in]{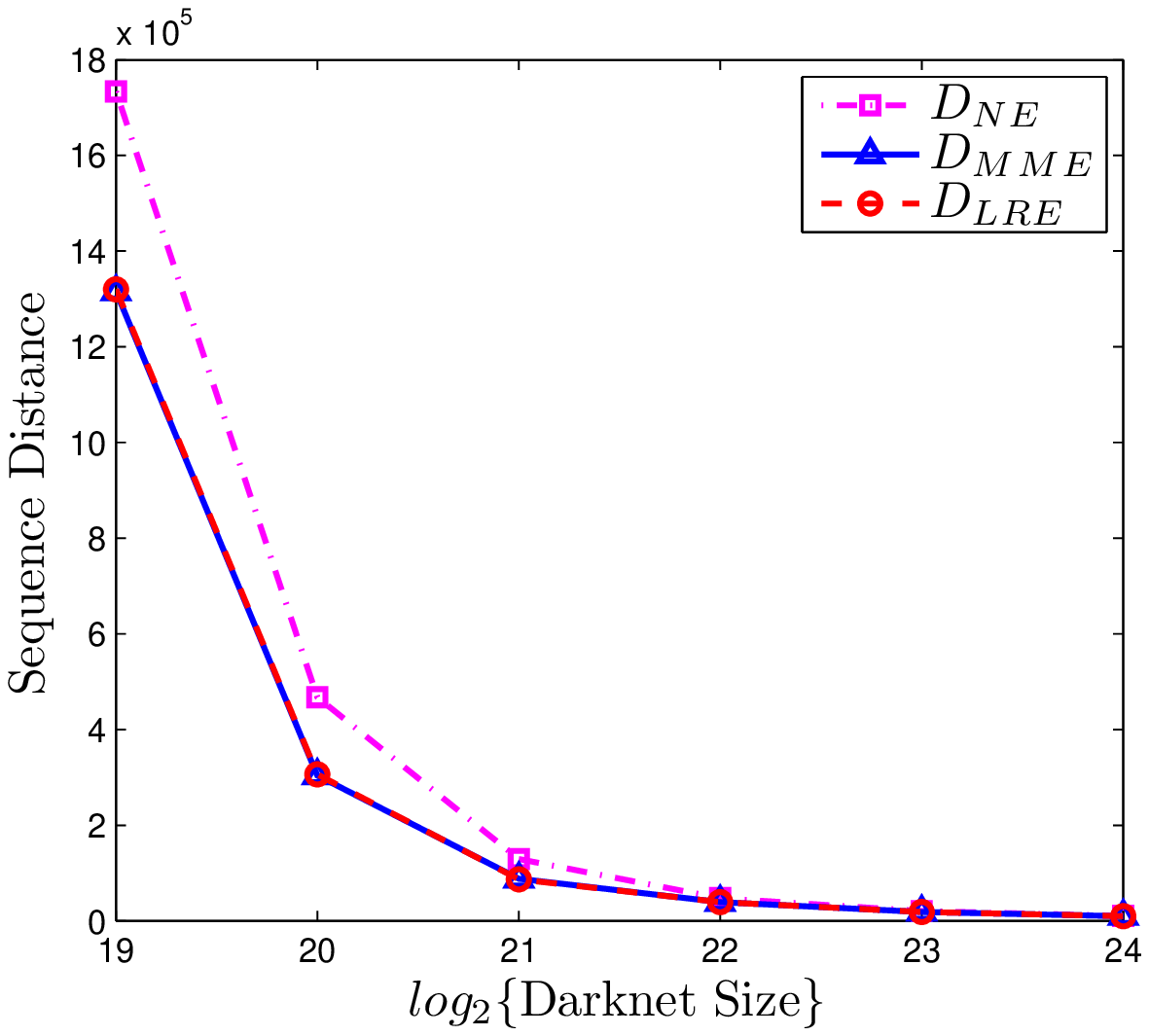}}
      \subfigure[]{\includegraphics[width=2.25in]{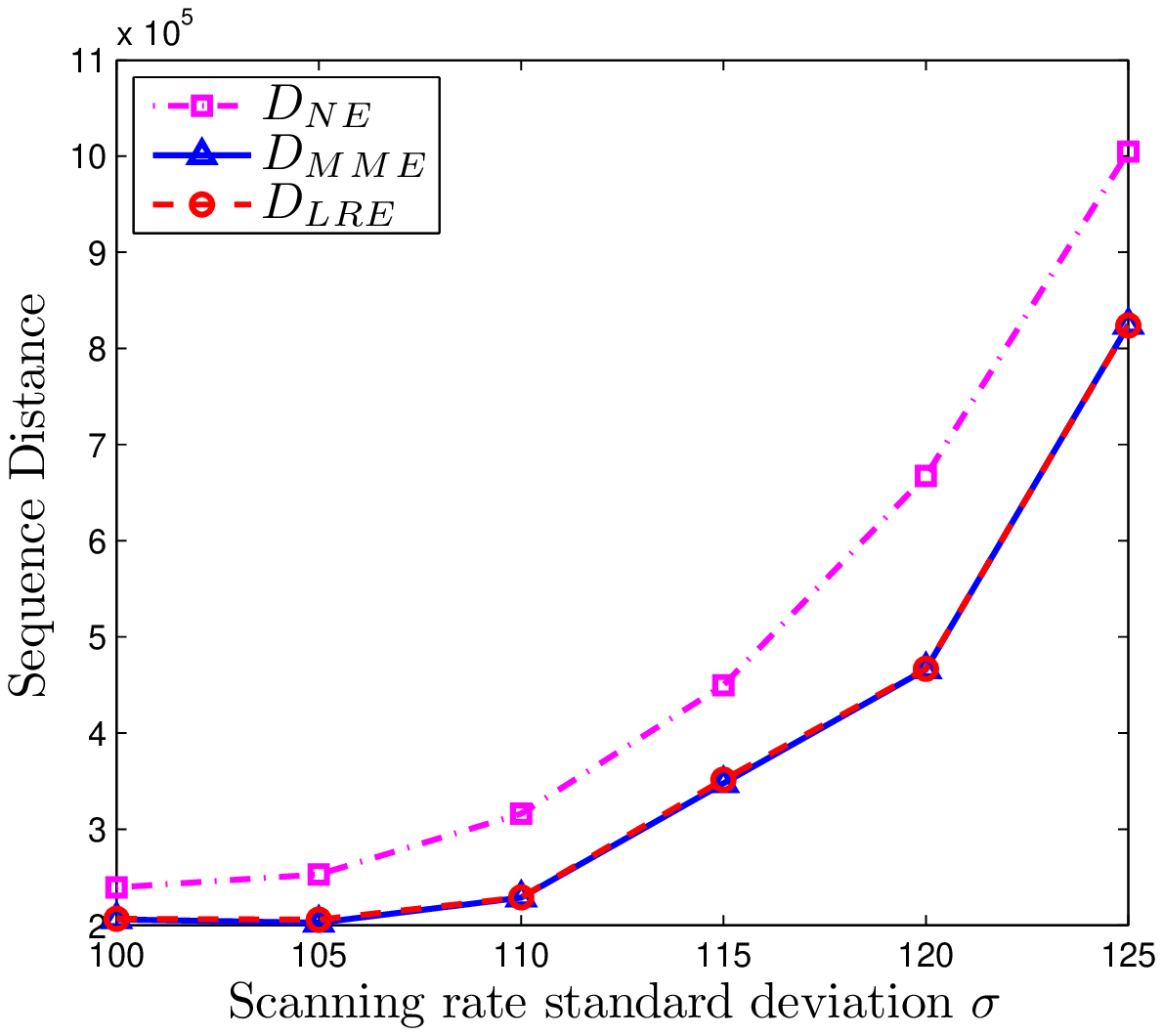}}
      \subfigure[]{\includegraphics[width=2.25in]{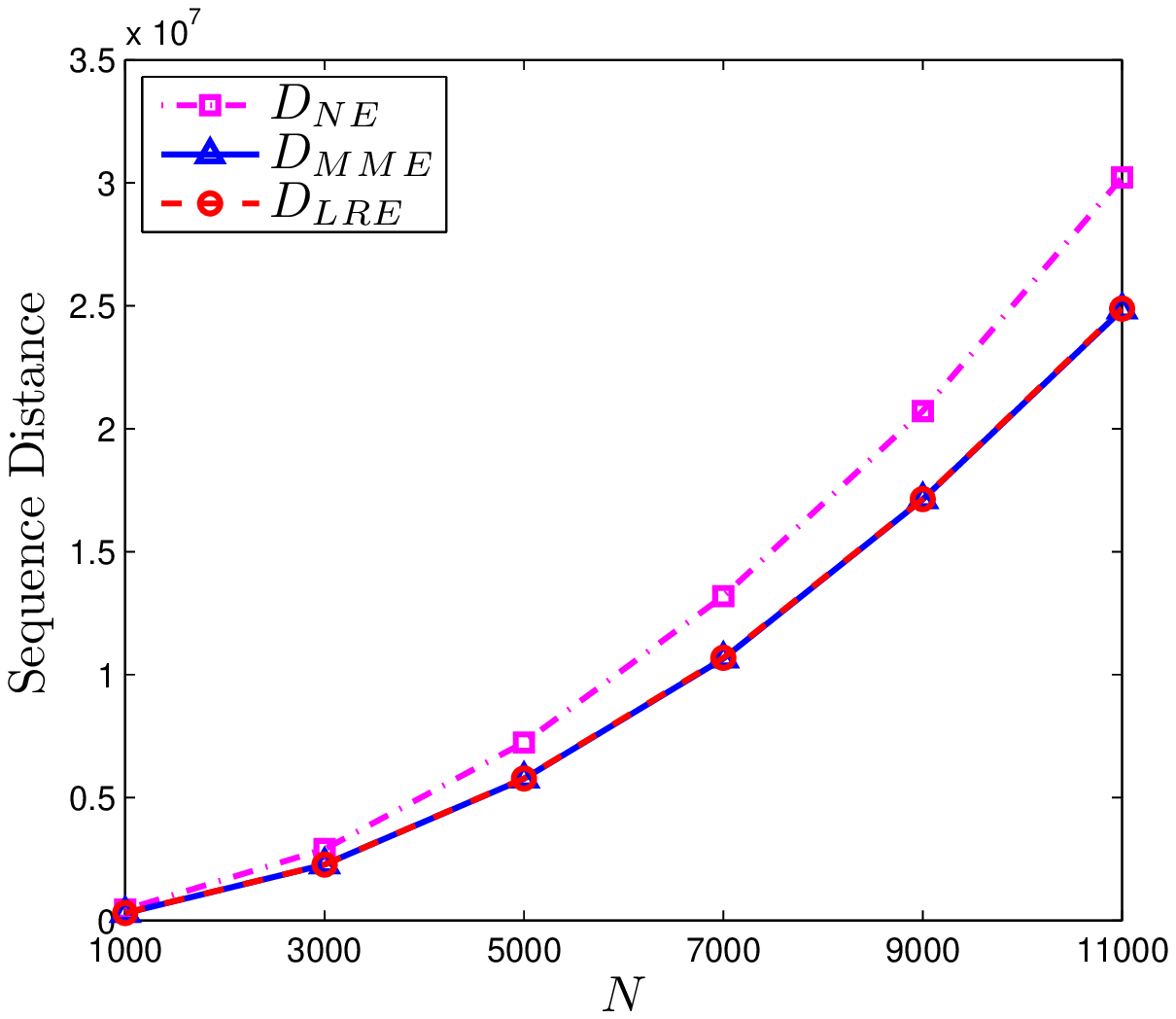}}
      }
        \caption{Simulation results of the sequence distance for random scanning.
    (a) Changing the Darknet size \big($N$ = 1,000, observation window size: 1,600 mins, scanning rate: $N(358, 115^2)$\big).
    (b) Changing the scanning rate standard deviation \big($N$ = 1,000, observation window size: 1,600 mins, Darknet size: $2^{20}$ IP addresses\big).
    (c) Changing the length of the infection sequence considered \big(observation window size: 1,600 mins, Darknet size: $2^{20}$ IP addresses, scanning rate: $N(358, 115^2)$\big).
    }
\label{seq}
\end{center}
\vspace{-0.3cm}
\end{figure*}
\begin{figure*}
\begin{center}
    \mbox{
    \hspace{-0.8cm}
      \subfigure[]{\includegraphics[width=2.25in]{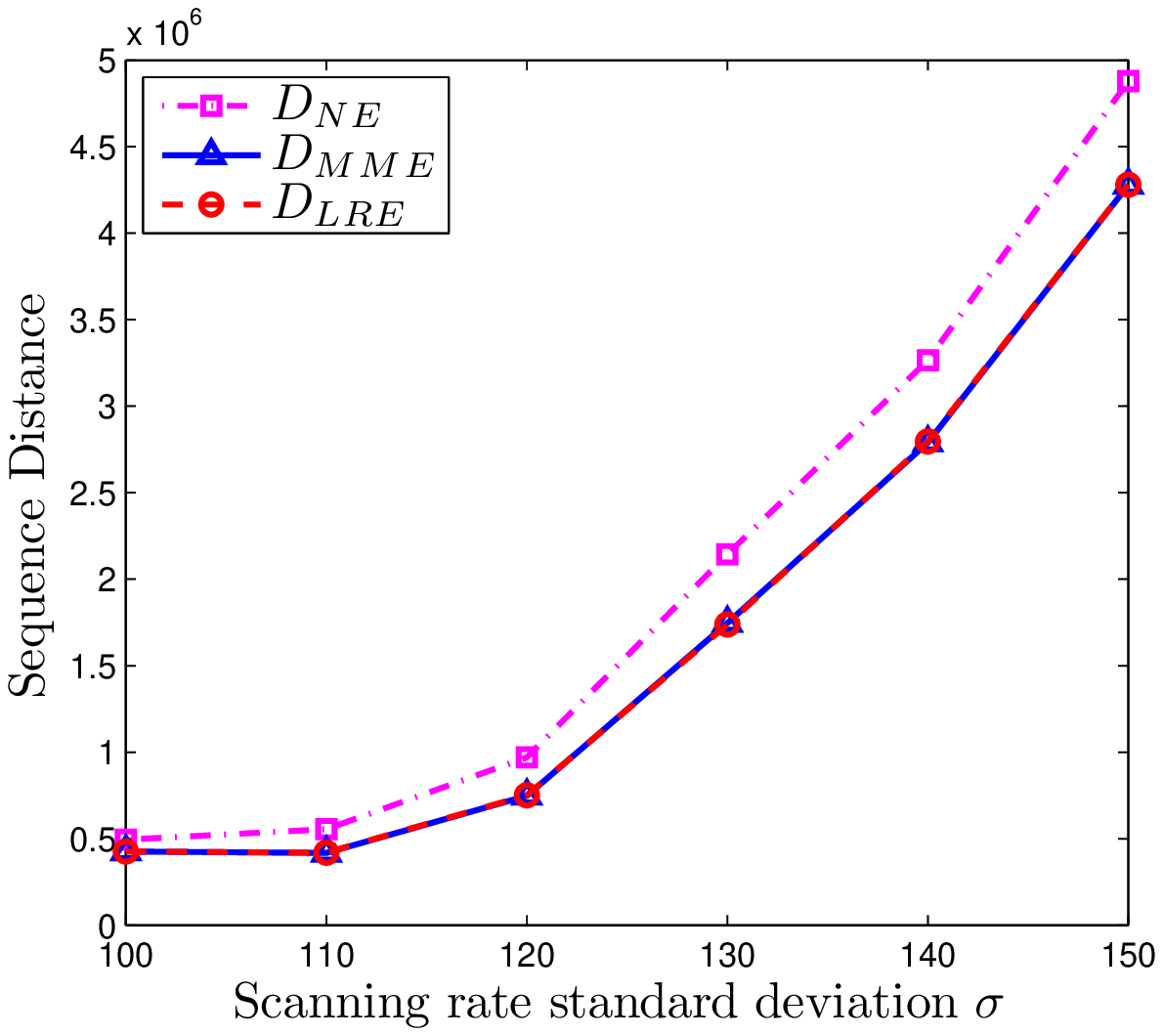}}
      \subfigure[]{\includegraphics[width=2.25in]{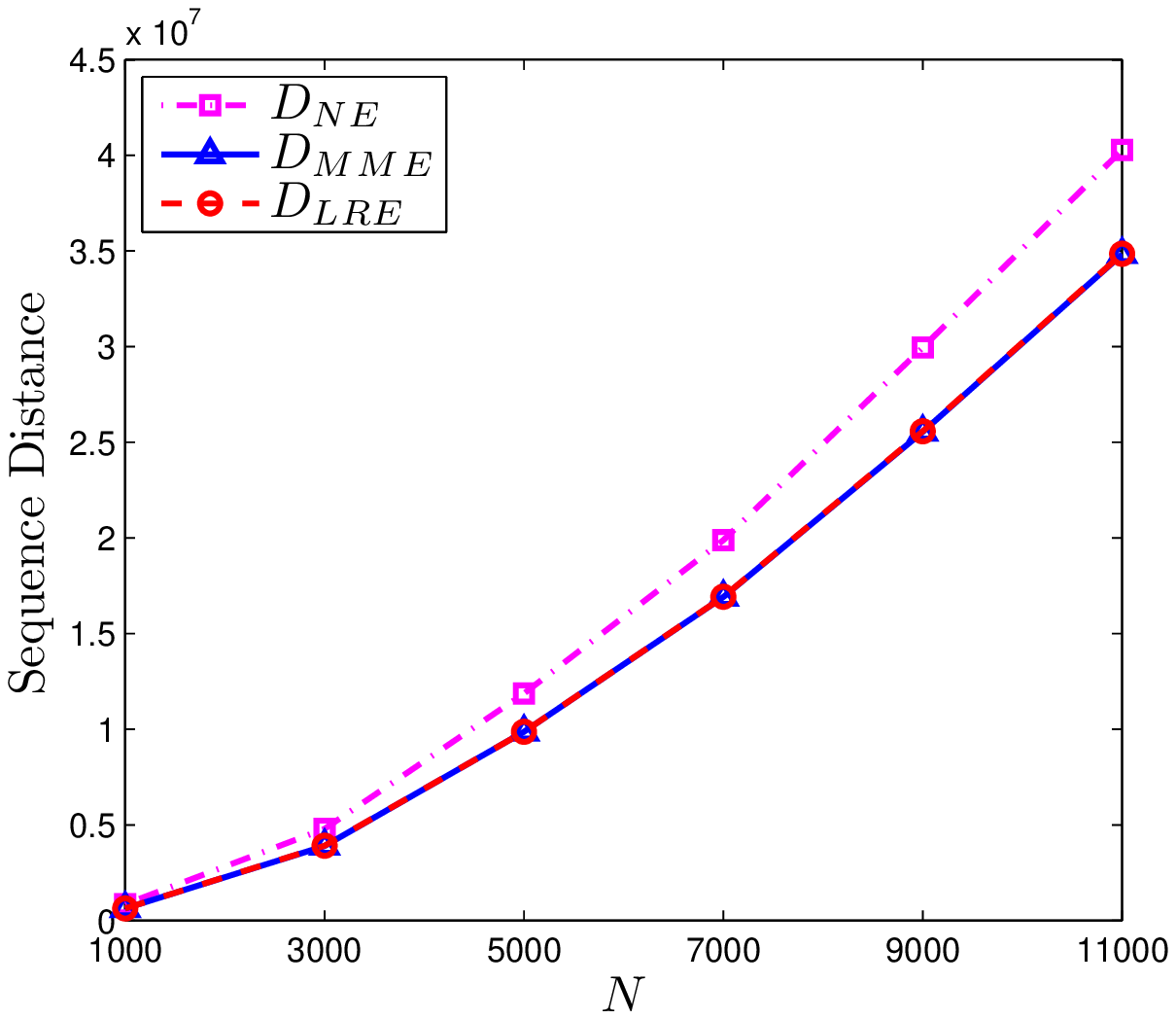}}
      \subfigure[]{\includegraphics[width=2.25in]{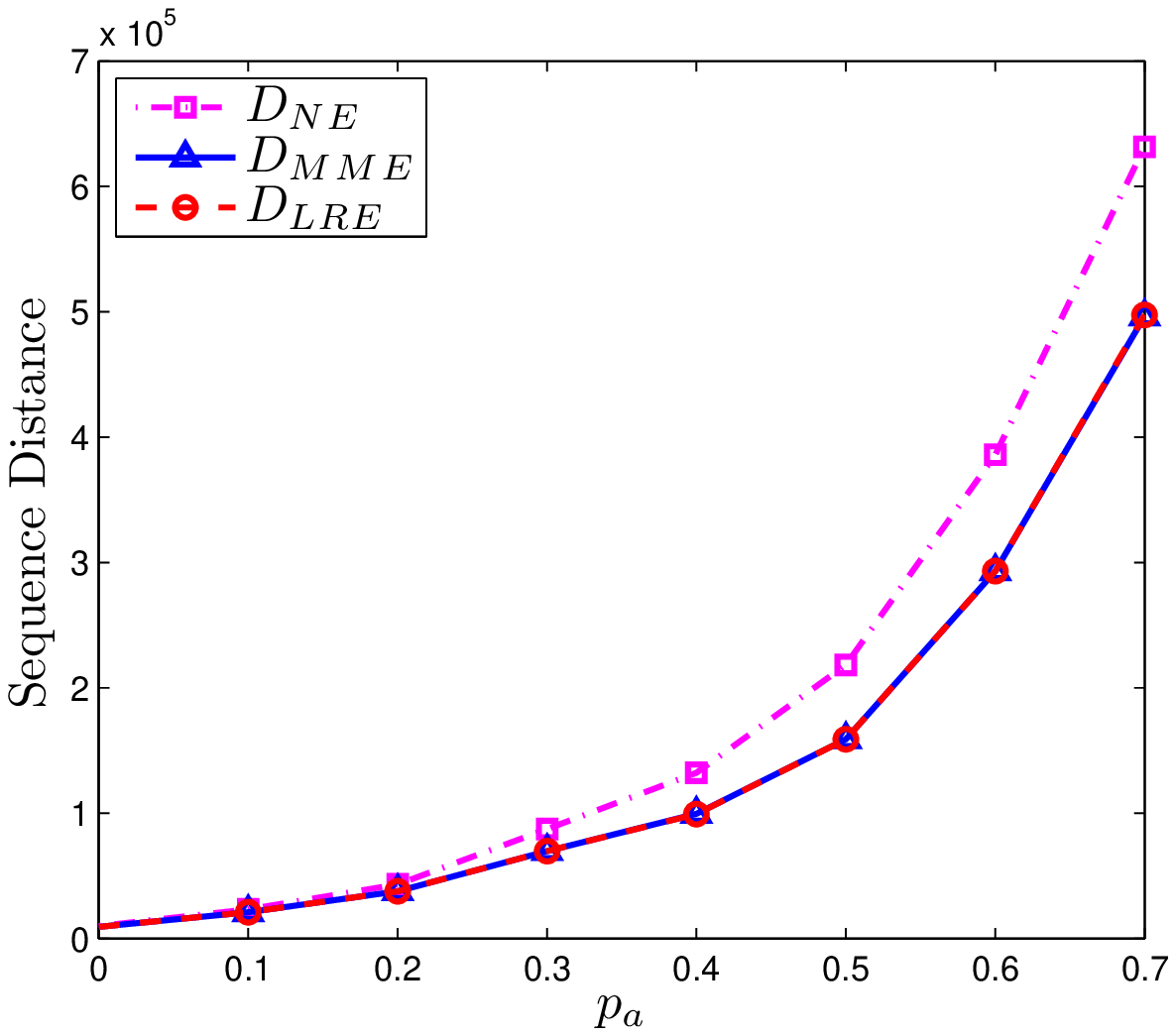}}
      }
        \caption{Simulation results of the sequence distance for localized scanning.
    (a) Changing the scanning rate standard deviation \big($p_a$ = 0.7, $N$ = 1,000, observation window size: 1,000 mins, Darknet size: $2^{24}$ IP addresses\big).
    (b) Changing the length of the infection sequence considered \big($p_a$ = 0.7, observation window size: 1,000 mins, Darknet size: $2^{24}$ IP addresses, scanning rate: $N(358, 115^2)$\big).
    (c) Changing the $p_a$ \big($N$ = 1,000, observation window size: 1,000 mins, Darknet size: $2^{24}$ IP addresses, scanning rate: $N(358, 115^2)$\big).
    }
\label{Lseq}
\end{center}
\vspace{-0.3cm}
\end{figure*}
\section{Simulation Results} \label{simulation}

In this section, we use simulations to verify our analytical results
and then apply estimators to identify the patient zero or the
hitlist. As far as we know, there is no publicly available data to show the real worm infection sequence.
That is, there is no dataset available with the real infection sequence to serve as the ground truth and a comparison basis for performance evaluation.
Therefore, we apply empirical simulations to provide the simulated worm infection time and infection sequence.

\subsection{Estimating the Host Infection Time}

We evaluate the performance of estimators in estimating the host
infection time. For the case of random-scanning worms, we simulate
the behavior of a host infected by the Code Red v2 worm. The host is
infected at time tick 0 and uses a constant scanning rate. The time
unit is set to 20 seconds. The Darknet records hit times during an
observation window. We consider the effects of the Darknet size, the
scanning rate, and the observation window size on the performance of
the estimators. The results are averaged over 100 independent runs.
Fig. \ref{size} compares the performance of NE, MME, and LRE with
different Darknet sizes from $2^{18}$ to $2^{25}$, a scanning rate
of 358 scans/min, and an observation window size of 800 mins. The
three sub-figures show the mean of estimators for $\mu$, the mean of
estimators for $t_0$, and the MSE of estimators for $t_0$. Fig.
\ref{rate} compares the three estimators with different scanning
rates from 158 scans/min to 558 scans/min, a Darknet size of
$2^{20}$, and an observation window size of 800 mins. Similarly,
Fig. \ref{win} is with different observation window sizes from 50
mins to 800 mins, a scanning rate of 358 scans/min, and a Darknet
size of $2^{20}$. It is observed that for all cases, our proposed
estimators have a better performance ({\em i.e.,} unbiasedness and
smaller MSE) than the naive estimator in estimating the host
infection time. Specifically, the simulation results verify Theorem
\ref{thm:time}, {\em i.e.,} that the MSE of our estimators is almost
half of that of the naive estimator, when the observation window
size is sufficiently large ({\em e.g.,} $>$ 200 mins).

Next, we study a host infected by localized-scanning worms and adopt
the same simulation parameters and settings as the above. The main
difference is that here the host preferentially searches for vulnerable
hosts in the ``local" address space with a probability $p_a$. In
Fig. \ref{fig:Lmse}, $p_a$ is set to 0.7, and we compare
MSE($\hat{t_0}$) for different estimators. We find that the results
are similar to those for the random-scanning case shown in Fig.s
\ref{size}-\ref{win}. The MSE($\hat{t_0}$) in Fig. \ref{fig:Lmse},
however, is larger for all cases since the localized-scanning worm
hits the Darknet less frequently than the random-scanning worm. In
Fig. \ref{Lpa}, we compare the performance of NE, MME, and LRE with
different $p_a$ from 0 to 0.9, a scanning rate of 358 scans/min, a
Darknet size of $2^{20}$, and an observation window size of 800
mins. Similarly, the results show that our estimators are unbiased
and the MSE of our estimators is almost half of that of the naive
estimator.

\begin{figure*}[tb]
\begin{center}
    \mbox{
      \subfigure[Random scanning.]{\includegraphics[width=2.5in]{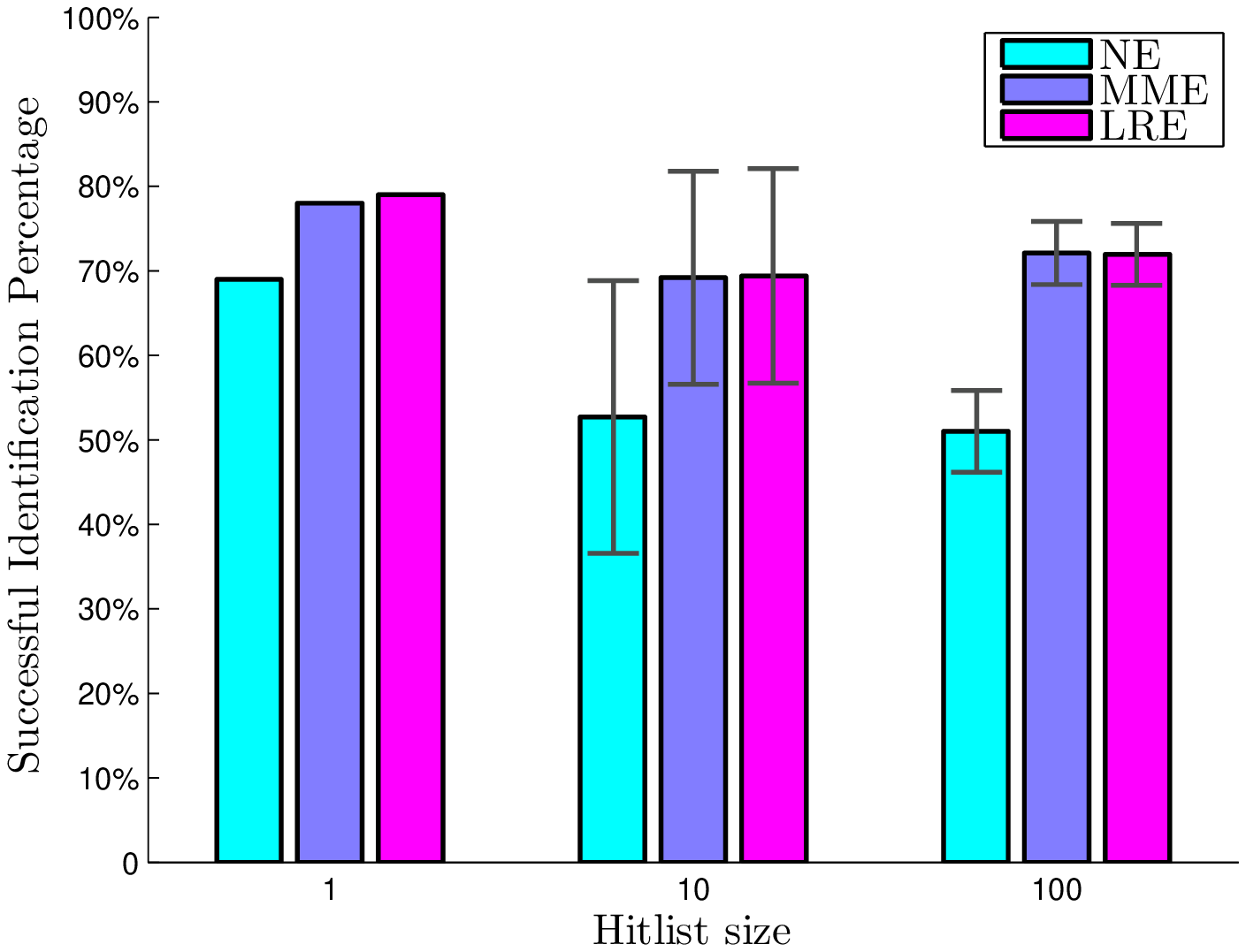}}
      \hspace{2cm}
       \subfigure[Localized scanning.]{\includegraphics[width=2.5in]{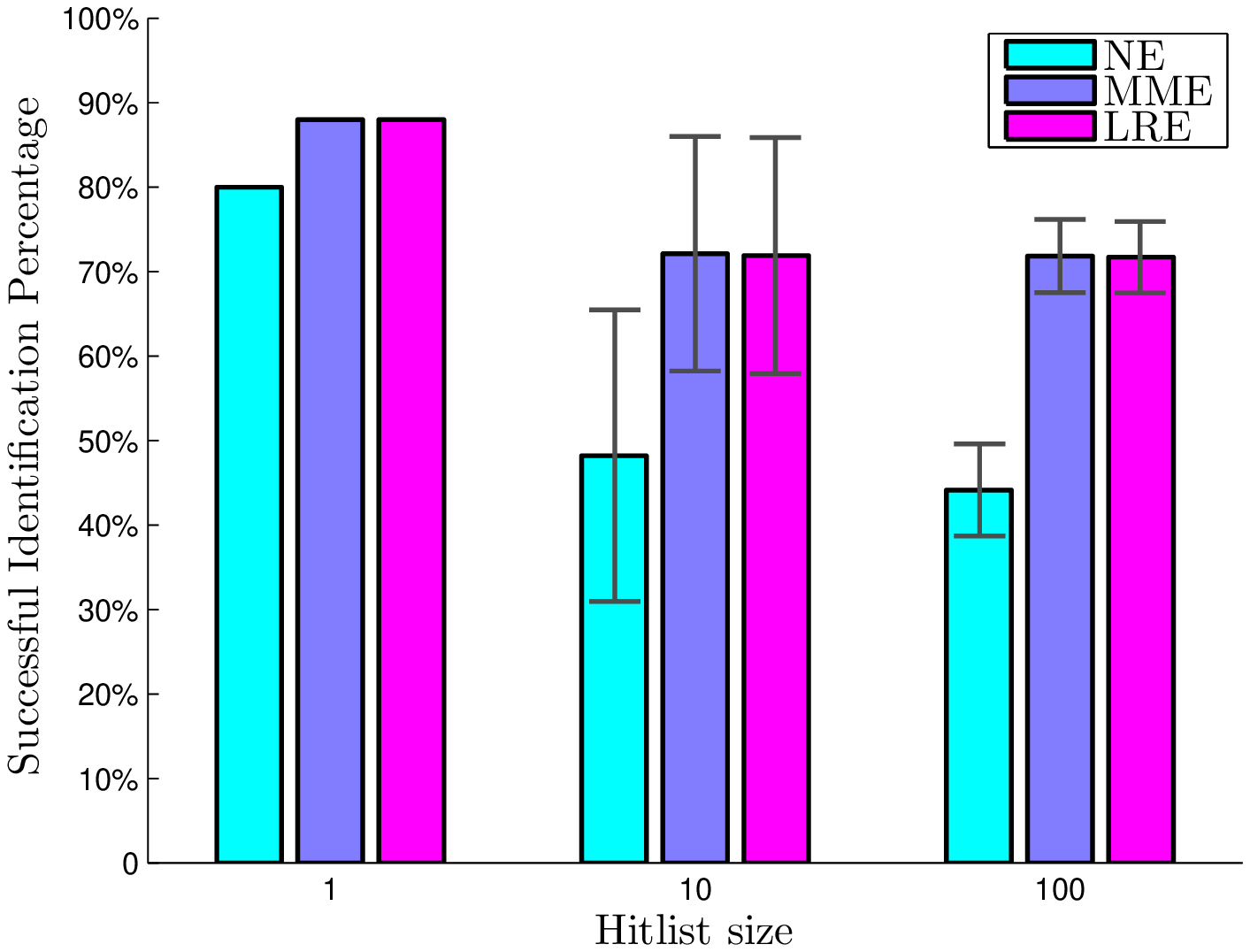}}
      }
          \caption{Comparison of estimators when changing the hitlist size. (a) Random scanning (all cases are for Darknet size: $2^{20}$ IP addresses, observation window size: 1000 mins,
    hitlist hosts scanning rate: $N(50, 20^2)$, other hosts scanning rate: $N(358, 110^2)$).
   (b) Localized scanning (all cases are for $p_a$ = 0.7, Darknet size: $2^{24}$ IP addresses, observation window size: 1000 mins,
    hitlist hosts scanning rate: $N(50, 20^2)$, other hosts scanning rate: $N(358, 110^2)$).}
   \label{fig:hitlist}
\end{center}
\vspace{-0.3cm}
\end{figure*}

\subsection{Estimating the Worm Infection Sequence}

We evaluate the performance of our algorithms in estimating the worm
infection sequence and simulate the propagation of the Code Red v2
worm. The simulator is extended from the code provided by
\cite{Simulator}, where the parameter setting is based
on the worm characteristics.
The Code Red worm has a vulnerable population of
360,000. Different infected hosts may have different scanning rates.
Thus, we assign a scanning rate (scans/min) from a normal
distribution $N(358, \sigma^2)$ to a newly infected host. Moreover,
we start our simulation at time tick 0 from one infected host. The
time unit is set to 20 seconds. Detailed information about how the
parameters are chosen can be found in Section VII of \cite{Zou05}.
Each point in Fig. \ref{seq} is
averaged over 20 independent runs.
Table \ref{tableseq} gives the results of a sample run with a Darknet size of $2^{20}$, an observation window size of 1,600 mins, and $\sigma$ = 110.
In the table, $S_i$ is the actual infection sequence ({\em i.e.,} $S_i=i$), whereas $\hat{S_i}$ is the estimated sequence.
In this example, we find that MME and LRE can pinpoint the patient zero successfully, while NE fails.

\begin{table}[tb]
\caption{A sample run of simulations for random scanning.} \vspace*{-10pt}
\def\temptablewidth{0.5\textwidth}
{\rule{\temptablewidth}{1pt}}
\begin{tabular*}{\temptablewidth}{@{\extracolsep{\fill}}cccccccc}
$S_i$& $\hat{S_i}_{\mbox {\tiny NE}}$& $\hat{S_i}_{\mbox {\tiny
MME}}$ & $\hat{S_i}_{\mbox {\tiny
LRE}}$ &$t_0$ &$\hat{t_0}_{\mbox {\tiny NE}}$& $\hat{t_0}_{\mbox {\tiny MME}}$& $\hat{t_0}_{\mbox {\tiny LRE}}$\\ 
\hline
1   &   2   &    1    &   1  &     0     &    114     &  20     &  20     \\
2   &   1   &    2    &   2  &     85    &     98     &   74    &   73 \\
3   &   3   &    3    &   3  &     105   &     165    &   116   &   116   \\
: & : &: &: &: &:&:&:\\
520   &  498    & 533  &   534  &   593 &    622   &  589    & 589   \\
521   &  433   &  488   &  477   &  594   &  611    & 581  &   580  \\
: & : &: &: &: &: &:&:\\
\end{tabular*}
       {\rule{\temptablewidth}{1pt}}
       \label{tableseq}
       \vspace{-0.3cm}
\end{table}

To compare the performance of estimators quantitatively, we consider
a simple $l_1$ {\em sequence distance}, {\em i.e.,}
    \begin{equation}
        D = \sum\limits_{i = 1}^N {\Big| {S_i  - \hat{S_i}} \Big|},
        \label{diff}
    \end{equation}
where $N$ is the length of the infection sequence considered. Note that the
smaller the sequence distance is, the better the estimator
performance will be. Fig. \ref{seq}(a) shows the sequence distances
of NE, MME, and LRE with varying Darknet sizes from $2^{19}$ to
$2^{24}$, an observation window size of 1,600 mins, $N$ = 1,000, and
$\sigma$ = 115. It is observed that when the Darknet size increases,
the performance of all estimators improves dramatically. Moreover,
the performance of MME and LRE is always better than that of NE. For
example, when the Darknet size equals $2^{19}$, MME and LRE improve
the inference accuracy by 24\%, compared with NE. Fig. \ref{seq}(b)
demonstrates the sequence distances of these three estimators by
changing the standard deviation of the scanning rate ({\em i.e.,}
$\sigma$) from 100 to 125. In the figure, the Darknet size is
$2^{20}$, the observation window size is 1,600 mins, and $N$ =
1,000. It is noted that when $\sigma$ increases, the performance of
all estimators deteriorates. The performance of MME and LRE,
however, is always better than that of NE. For example, when
$\sigma$ = 120, MME and LRE reduce the sequence distance by 30\%,
compared with NE. In Fig. \ref{seq}(c), we increase the length of
the infection sequence considered, $N$, from 1,000 to 11,000. Here
the Darknet size is $2^{20}$, the observation window size is 1,600
mins, and $\sigma$ = 115. It is intuitive that the sequence
distances of all estimators become larger as $N$ increases. However,
MME and LRE are always better than NE.

Next, we extend our simulator to imitate the spread of
localized-scanning worms. Specifically, we consider /8
localized-scanning worms and a centralized /8 Darknet with $2^{24}$ IP addresses.
We still use the Code Red v2 worm parameters and the same setting as
random scanning, except that the observation window size is 1,000
mins (this is because localized-scanning worms spread faster). The distribution of vulnerable hosts is extracted from the
dataset provided by DShield \cite{DShield}. DShield obtains the
information of vulnerable hosts by aggregating logs from more than
1,600 intrusion detection systems distributed throughout the
Internet. Specifically, we use the dataset with port 80 (HTTP) that
is exploited by the Code Red v2 worm to generate the
vulnerable-hosts distribution. Each point in Fig. \ref{Lseq} is
averaged over 20 independent runs. Fig. \ref{Lseq} compares the
sequence distances of different estimators for localized scanning.
Specifically, the results in Fig. \ref{Lseq}(a) and (b) are similar
to those in Fig. \ref{seq}(b) and (c). In Fig. \ref{Lseq}(c), we
compare the performance of the estimators by increasing $p_a$ from 0
to 0.7. Here, $N$ = 1,000, and $\sigma$ = 115. It is observed that
the sequence distances of all estimators increase as $p_a$ becomes
larger. However, our estimators are always better than NE. For
example, when $p_a$ = 0.5, MME and LRE increase the inference
accuracy by 27\%, compared with NE.

Therefore, our proposed
estimators perform much better than the naive estimator for both
random-scanning and localized-scanning worms in estimating the worm infection sequence.

\subsection{Identifying the Patient Zero or the Hitlist}

As discussed in Section 1, a smart worm can assign lower
scanning rates to the initially infected host(s) and higher scanning
rates to other infected hosts. In this way, the Darknet might observe
later infected hosts first, and therefore the smart worm would weaken
the performance of the naive estimator.
In Fig. \ref{fig:hitlist}, we
compare the performance of estimators in identifying the hitlist of
such a smart worm. Specifically, the worm assigns scanning rates
from $N(50, 20^2)$ to the host(s) on the hitlist and scanning rates
from $N(358, 110^2)$ to other infected hosts. Then, we calculate the
percentage of the host(s) on the hitlist that are successfully
identified by an estimator. For example, if the size of the hitlist
is 100 and 50 hosts that belong to the hitlist are identified among
the first 100 hosts of the estimated infection sequence, the
successful identification percentage of the estimator is 50\%. The
results are averaged over 100 independent runs. Fig.
\ref{fig:hitlist}(a) shows the case of random scanning, where the
Darkent size is $2^{20}$ and the observation window size is 1,000
mins. It is seen that our estimators have a higher successful
identification percentage and a smaller variance than the naive
estimator. For instance, when the size of the hitlist is 1 ({\em
i.e.,} the worm starts from the patient zero), MME and LRE can
pinpoint the patient zero around 80\% of the time, while NE can
detect it only 70\% of the time. When the size of the hitlist is 10
or 100, compared with NE, our proposed estimators increase the
number of successfully identified hosts from 5 to 7 or 51 to 72, and
reduce the variance from 2.6 to 1.6 or 23 to 13, respectively. Fig.
\ref{fig:hitlist}(b) shows the results of localized scanning, where
the Darkent size is $2^{24}$ and $p_a$ = 0.7, and all other
parameters are the same as the case of random scanning. The results
are similar to those in Fig. \ref{fig:hitlist}(a). Therefore, the
simulation results demonstrate that our proposed estimators are much
more effective in identifying the histlist of the smart worm than
the naive estimator.

\section{Discussions}\label{discussion}

In this section, we first analyze the chance that Darknet misses an
infected host and then discuss the limitations and the extensions of
our proposed estimators.

\subsection{Host Missing Probability}

By applying Darknet observations, we have made an assumption: The
infected host will hit the Darknet. Then, an intuitive question would be: What is the
probability that the Darknet misses an infected host within a given
observation window?

We consider the case of localized scanning and regard random
scanning as a special case of localized scanning when $p_a$ = 0. The
probability for a scan from an infected host to hit the Darknet is
$(1-p_a)\cdot \omega/\Omega$; and then the probability that the
Darknet misses observing the host in a time unit is $(1-
(1-p_a)\cdot{\omega}/{\Omega})^s$. Thus, the host missing
probability ({\it i.e.}, the probability that the Darknet misses the
infected host in a $k$ time units observation window) is
    \begin{equation}
    \label{equ:p_rs_miss}
         \mbox{Pr}_{\mbox {\tiny LS}} (\mbox{missing}) = \Big(1- (1-p_a)\cdot \frac{\omega}{\Omega}\Big)^{s\cdot k}.
    \end{equation}

\begin{figure}[tb]
\begin{center}
\includegraphics [width=2.7in]{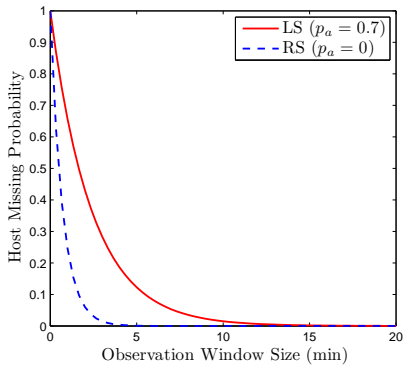}
\caption{Host missing probability ($p_a$ = 0.7, Darknet size: $2^{24}$ IP addresses, scanning rate: 358 scans/min).}
\label{missing}
\end{center}
\vspace{-0.3cm}
\end{figure}

In Fig. \ref{missing}, we show the host missing probability as the
observation window size changes. In this example, we set $\omega$ =
$2^{24}$, time unit = 20 seconds, and $s$ = 358 scans/min. We find that if $p_a$ = 0.7, the infected host will almost
hit the Darknet for sure when the observation window size is larger
than 20 mins. If $p_a$ = 0, which is the case of random scanning,
a 5-min observation window is sufficient to guarantee the capture of
the infected host. Therefore, in our previous analysis and
simulation, the assumption that the Darknet can observe scans from the infected
host, especially at the early stage, is reasonable. Moreover, our
estimator can still work even for self-stopping worms \cite{Ma}.

\subsection{Estimator Limitations and Extensions}

Our proposed estimators are built based on some assumptions listed
in Section \ref{time}. Attackers that design future worms may exploit these assumptions to
weaken the accuracy of our estimators. In the following, we discuss
some limitations of our estimators and the potential extensions.
\subsubsection{Darknet Avoidance}

The majority of active worms up to date do not attempt to avoid the
detection of Darknet. As a result, CAIDA's network telescopes have
been observing many active Internet worms such as Code Red, Slammer,
Witty, and even recently the Conficker worm (also known as the April
Fool's worm). Most worms apply random scanning and localized
scanning, and Darknet can observe the traffic from such worms.

Recent work, however, has shown that attackers can potentially detect the
locations of Darknet or network sensors \cite{Bethencourt}. Thus, a
future worm can be specially designed to avoid scanning the address space of the
Darknet. The countermeasure against such an intelligent worm is to
apply the distributed Darknet instead of the centralized Darknet
\cite{Moore04}. That is, unused IP addresses in many subnets are
used to observe worm traffic, which is then reported to a collection
center for further processing.
A prototype of
distributed Darknet has been designed and evaluated in \cite{Bailey}.

\subsubsection{Scanning Rate Variation}

Although there have been no observations of worms that use scanning rate variation mechanisms
(\textit{i.e.,} the scanning rate of an individual infected host is time-variant) \cite{Wei},
future worms may employ such schemes to invalidate our
basic assumption and thus weaken the performance of our estimators. Changing the scanning rate, however, introduces
additional complexity to worm design and can slow down worm
spreading. Moreover, if the change of scanning rates is relatively
slow, our estimators can be enhanced with the change-point detection
\cite{Basseville} to detect and track when the scanning rate has a significant
change and then apply the early observations to derive the infection
time of an infected host.

\begin{figure}[tb]
\begin{center}
\includegraphics [width=2.7in]{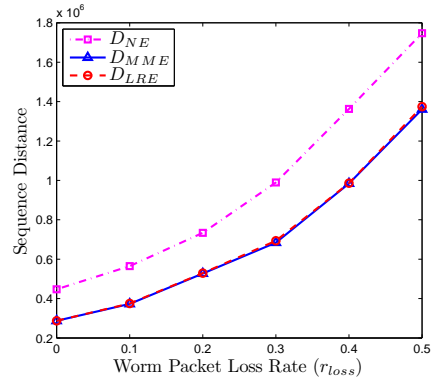}
\caption{Simulation results of the sequence distances of different estimators varying with the worm packet loss rate.
 ($N$ = 1,000, observation window size: 1,600 mins, scanning rate: $N(358, 115^2)$, Darknet size: $2^{20}$ IP addresses).
 }
\label{fig:loss}
\end{center}
\vspace{-0.3cm}
\end{figure}
\subsubsection{Measurement Errors}
The measurement errors can affect the performance of
estimators. There are two types of measurement errors. The false positive denotes that Darknet incorrectly classifies
the traffic from a benign host as worm traffic, whereas
the false negative is that Darknet incorrectly classifies worm
traffic as benign traffic or misses worm traffic due to
congestion or device malfunction.

For the false positives, most of time we can distinguish worm
traffic from other traffic. First, our estimation techniques are
used as a form of post-mortem analysis on worm records logged by
Darknet. As a result, we can limit our analysis to the records
logged during the outbreak of the worm when it is most rampant. More
importantly, worm packages always contain information about infection vectors that
distinguish worm traffic from other traffic. For example, the Witty
worm uses a source port of 4,000 to attack Internet Security Systems
firewall products [16]. It is very unlikely that a benign host uses
a source port of 4,000. By filtering the records based on infection
vectors specific to the worm under investigation, we can eliminate
most of the effects of false positives on Darknet observations.

False negatives are much harder to eliminate. A packet towards
Darknet may be lost due to congestion caused by the worm (such as the
Slammer worm \cite{Moore03}) or the
malfunction of Darknet monitoring devices. To study the effects of
false negatives, we modify our simulator to mimic the packet loss
and evaluate the performance of our estimators under false
negatives. Here we assume that the loss rate of the worm packets towards Darknet (denoted as $r_{\mbox{\scriptsize loss}}$) is
the same for each infected host. Fig. \ref{fig:loss} shows how the sequence
distances of different estimators vary with the worm packet loss
rate. The results are averaged over 20 independent runs. It is intuitive that when the packet loss rate becomes larger,
the performance of all estimators worsens. Our proposed estimators,
however, always perform much better than NE.
For example, compared with NE, our estimators ({\em i.e.,} MME and LRE) improve the inference accuracy by
28\% when $r_{\mbox{\scriptsize loss}}$ = 0.4.
A mechanism to recover from worm-induced
congestion has been proposed in \cite{Wei}, which estimates the
packet loss rates of infected hosts based on Darknet observations
and BGP atoms. This method can be incorporated into our estimators
to enhance their robustness against worm-induced congestion.

\section{Related Work} \label{work}

Under the framework of Internet worm tomography, several works have
applied Darknet observations to infer the characteristics of worms.
For example, Chen {\em et al.} studied how the Darknet can be used
to monitor, detect, and defend against Internet worms \cite{AAWP}.
Moore {\em et al.} applied network telescope observations and least
squares fitting methods to infer the number of infected hosts and
scanning rates of infected hosts \cite{Moore04}. Some works have
researched on how to use Darknet observations to detect the
appearance of worms \cite{Wu04,Zou05,Bu06,Soltani08}. For instance,
Zou {\em et al.} used a Kalman filter to infer the infection rate of
a worm and then detect the worm \cite{Zou05}. Moreover, the Darknet
observations have been used to study the feature of a specific worm, such
as Code Red \cite{Moore02}, Slammer \cite{Moore03}, and Witty
\cite{Shannon}.

Internet worm tomography has been applied to infer worm temporal
behaviors. For example, Kumar {\em et al.} used network telescope
data and analyzed the pseudo-random number generator to reconstruct
the ``who infected whom" infection tree of the Witty worm
\cite{Kumar05}. Hamadeh {\em et al.} further described a general
framework to recover the infection sequence for both TCP and UDP
scanning worms from network telescope data \cite{Hamadeh06}. Rajab
{\em et al.} applied the same data and studied the ``infection and
detection times" to infer the worm infection sequence
\cite{Rajab05}. Different from the above works, in this work we
employ advanced statistical estimation techniques to Internet worm
tomography.

\section{Conclusions}\label{conclude}
In this paper, we have attempted to understand the temporal
characteristics of Internet worms through both analysis and
simulation under the framework of Internet worm tomography.
Specifically, we have proposed method of moments, maximum
likelihood, and linear regression estimators to infer the host
infection time and reconstruct the worm infection sequence. We have
shown analytically and empirically that the mean squared error of
our proposed estimators can be almost half of that of the naive
estimator in estimating the host infection time. Moreover, we have
formulated the problem of estimating the worm infection sequence as
a detection problem and have calculated the probability of error
detection for different estimators. We have demonstrated empirically
that our estimation techniques perform much better than the
algorithm used in \cite{Rajab05} in estimating the worm infection
sequence and in identifying the hitlist for both random-scanning and
localized-scanning worms.


%

\appendices

\renewcommand{\theequation}{\thesection.\arabic{equation}}

\section{Table \ref{tab:mu}: Estimator Properties ($\hat{\mu}$)}

We calculate the bias, the variance, and the MSE of different
estimators for estimating $\mu$.

\subsection{Naive Estimator}

Since $\hat{\mu}_{\mbox{\tiny NE}}=1$, the bias of NE is
\begin{equation}
\mbox{Bias}(\hat{\mu}_{\mbox {\tiny NE}}) =
\mbox{E}(\hat{\mu}_{\mbox {\tiny NE}})-\mu = 1 - \textstyle {1\over
p}.
\end{equation}

\noindent Note that $\hat{\mu}_{\mbox{\tiny NE}}$ is constant. Thus,
the variance of NE is
\begin{equation}
\mbox{Var}(\hat{\mu}_{\mbox {\tiny NE}}) = \mbox{E}\,[(\hat{\mu}_{\mbox {\tiny NE}}-\mbox{E}(\hat{\mu}_{\mbox {\tiny NE}}))^2]=0.
\end{equation}

\noindent Therefore,
\begin{equation}
\mbox{MSE}(\hat{\mu}_{\mbox {\tiny NE}}) =
\mbox{Bias}^2(\hat{\mu}_{\mbox {\tiny
NE}})+\mbox{Var}(\hat{\mu}_{\mbox {\tiny NE}}) = \textstyle
\frac{(1-p)^2}{p^2}.
\end{equation}

\subsection{Method of Moments Estimator / Maximum Likelihood Estimator}

Since $\mbox{E}(\delta_i)=\mu$ for $i=1,2,\cdots,n-1$ and Equations
(\ref{equ:MME_u}) and (\ref{equ:MLE_u}) hold, the bias of
$\hat{\mu}_{\mbox {\tiny MME}}$ (or $\hat{\mu}_{\mbox {\tiny MLE}}$)
is calculated as
\begin{equation}
\mbox{Bias}(\hat{\mu}_{\mbox {\tiny MME}})=\textstyle
\mbox{E}\Big({1 \over n-1}{\sum\limits_{i = 1}^{n -
1}{\delta_i}}\Big)-\mu =0,
\end{equation}
which is unbiased. Note that $\mbox{Var}(\delta_i)=\frac{1-p}{p^2}$
for $i=1,2,\cdots,n-1$ and $\delta_i$'s are independent. Thus, we
have
\begin{equation}
\label{equ:Var_MME_u}
\mbox{Var}(\hat{\mu}_{\mbox {\tiny MME}}) =
\textstyle \mbox{Var}\Big({1 \over n-1}{\sum\limits_{i = 1}^{n -
1}{\delta_i}}\Big) = \textstyle \frac{1-p}{p^2(n-1)}.
\end{equation}

\noindent Therefore, the MSE of $\hat{\mu}_{\mbox {\tiny MME}}$ (or
$\hat{\mu}_{\mbox {\tiny MLE}}$) is
\begin{equation}
\mbox{MSE}(\hat{\mu}_{\mbox {\tiny MME}}) =
\mbox{Bias}^2(\hat{\mu}_{\mbox {\tiny
MME}})+\mbox{Var}(\hat{\mu}_{\mbox {\tiny MME}}) = \textstyle
\frac{1-p}{p^2(n-1)}.
\end{equation}
It is noted that for an unbiased estimator, the MSE is identical to
its variance.

\subsection{Linear Regression Estimator}

Note that $\hat{\mu}_{\mbox {\tiny LRE}}=\frac{\overline {i \cdot
t}- \overline i \cdot \overline t}{\overline {i^2} - (\overline i)
^2}$. From Equation (\ref{equ:LRE_1}) and
$t_i=t_0+\sum_{j=0}^{i-1}{\delta_j}$, $i=1,2,\cdots,n$, we have
\begin{eqnarray}
 \overline {i\cdot t} &=& \frac{1}{n}\sum\limits_{i = 1}^n i \cdot t_i \nonumber\\
                      &=& \frac{n+1}{2}t_0+\frac{1}{n}\sum_{i=0}^{n-1} {\sum_{j=i+1}^n {j\cdot\delta_i}} \nonumber \\
                      &=& \frac{n+1}{2}t_0+\sum_{i=0}^{n-1}{\frac{(n-i)(n+i+1)}{2n}\delta_i}
\end{eqnarray}
and
\begin{equation}
    \overline{i} \cdot \overline{t} = \overline{i}\cdot \frac{1}{n}\sum_{i=1}^{n}{t_i} =\overline{i}\cdot
    t_0+\overline{i}\cdot \sum_{i=0}^{n-1}{\frac{n-i}{n}\delta_i}.
\end{equation}

\noindent Since $\overline{i}=\frac{n+1}{2}$ and
$\overline{i^2}=\frac{(n+1)(2n+1)}{6}$,
\begin{equation}
\label{equ:1}
    \overline{i\cdot t}-\overline{i}\cdot \overline{t} =
    \sum_{i=1}^{n-1}{\frac{i(n-i)}{2n}\delta_i}
\end{equation}
and
\begin{equation}
\label{equ:2}
    \overline{i^2}-\left(\overline{i} \right)^2 = \frac{n^2-1}{12}.
\end{equation}

\noindent Note that $\mbox{E}(\delta_i)=\mu$ and
$\mbox{Var}(\delta_i)=\frac{1-p}{p^2}$, $i=0,1,\cdots,n-1$, and
$\delta_i$'s are independent. Moreover,
$\sum_{i=1}^{n}{i^3}=\left(\frac{n(n+1)}{2}\right)^2$ and
$\sum_{i=1}^{n}{i^4}=\frac{1}{30}(6n^5+15n^4+10n^3-n)$. Then, we
have
\begin{equation}
   \mbox{E}(\overline{i\cdot t}-\overline{i}\cdot \overline{t}) =
   \sum_{i=1}^{n-1}{\frac{i(n-i)}{2n}\mu} = \frac{n^2-1}{12}\mu
\end{equation}
and

\begin{eqnarray}
   \mbox{Var}(\overline{i\cdot t}-\overline{i}\cdot \overline{t})
   &=& \sum_{i=1}^{n-1}{\left(\frac{i(n-i)}{2n}\right)^2\cdot \frac{1-p}{p^2}} \nonumber \\
   &=& \frac{1-p}{4n^2p^2}\left(n^2\sum_{i=1}^{n-1}{i^2}-2n\sum_{i=1}^{n-1}{i^3}+\sum_{i=1}^{n-1}{i^4}\right) \nonumber \\
   &=& \frac{1-p}{p^2}\cdot \frac{n^4-1}{120n}.
\end{eqnarray}

\noindent Therefore, the bias of $\hat{\mu}_{\mbox {\tiny LRE}}$ can
be calculated as
\begin{equation}
\mbox{Bias}(\hat{\mu}_{\mbox {\tiny LRE}}) = \textstyle
\mbox{E}\Big(\frac{\overline {i \cdot t}- \overline i \cdot
\overline t}{\overline {i^2} - (\overline i) ^2}\Big)-\mu =0,
\end{equation}
which is unbiased. Moreover, the variance and the MSE of
$\hat{\mu}_{\mbox {\tiny LRE}}$ are
\begin{eqnarray}
\label{equ:Var_LRE_u}
 \mbox{MSE}(\hat{\mu}_{\mbox {\tiny LRE}}) &=& \mbox{Var}(\hat{\mu}_{\mbox {\tiny LRE}})\nonumber \\
                                           &=& \textstyle \mbox{Var}\Big(\frac{\overline {i \cdot t}- \overline i \cdot
\overline t}{\overline {i^2} - (\overline i) ^2}\Big) \nonumber \\
                                           &=& \textstyle \frac{6(n^2+1)(1-p)}{5n(n^2-1)p^2}.
\end{eqnarray}

\section{Table \ref{tab:t0}: Estimator Properties ($\hat{t_0}$)}

We calculate the bias, the variance, and the MSE of different
estimators for estimating $t_0$.

\subsection{Naive Estimator}

Since $\hat{t_0}_{\mbox {\tiny NE}} = t_1-\hat{\mu}_{\mbox {\tiny
NE}}=t_0+\delta_0-1$, $\mbox{E}(\delta_0)=\frac{1}{p}$, and
$\mbox{Var}(\delta_0)=\frac{1-p}{p^2}$,
\begin{eqnarray}
\mbox{Bias}(\hat{t_0}_{\mbox {\tiny NE}})& =&
t_0+\mbox{E}(\delta_0)-1-t_0 = \textstyle \frac{1-p}{p}\\
\mbox{Var}(\hat{t_0}_{\mbox {\tiny NE}}) &=&
\mbox{Var}(t_0+\delta_0-1) = \textstyle  \frac{1-p}{p^2}\\
\mbox{MSE}(\hat{t_0}_{\mbox {\tiny
NE}})&=&\mbox{Bias}^2(\hat{t_0}_{\mbox {\tiny
NE}})+\mbox{Var}(\hat{t_0}_{\mbox {\tiny NE}}) \nonumber \\
&=& \textstyle
\frac{(1-p)(2-p)}{p^2}.
\end{eqnarray}
Note that when $p\ll 1$, $\mbox{MSE}(\hat{t_0}_{\mbox {\tiny NE}})
\approx \textstyle \frac{2(1-p)}{p^2}$.

\subsection{Method of Moments Estimator / Maximum Likelihood Estimator}

Note that $\hat{t_0}_{\mbox {\tiny MME}}=\hat{t_0}_{\mbox {\tiny
MLE}}=t_0+\delta_0-\hat{\mu}_{\mbox {\tiny MME}}$ and
$\mbox{E}(\delta_0)=\mbox{E}(\hat{\mu}_{\mbox {\tiny MME}})=\mu$.
Thus,
\begin{equation}
\mbox{Bias}(\hat{t_0}_{\mbox {\tiny MME}}) =
t_0+\mbox{E}(\delta_0)-\mbox{E}(\hat{\mu}_{\mbox {\tiny MME}})-t_0 =
0
\end{equation}
\begin{equation}
\mbox{MSE}(\hat{t_0}_{\mbox {\tiny MME}})
=\mbox{Var}(\hat{t_0}_{\mbox {\tiny MME}}) =
\mbox{Var}(\delta_0-\hat{\mu}_{\mbox {\tiny MME}}).
\end{equation}

\noindent Since $\hat{\mu}_{\mbox {\tiny MME}} = {1 \over
n-1}{\sum_{i=1}^{n-1}{\delta_i}}$ that is independent of $\delta_0$,
\begin{eqnarray}
\mbox{MSE}(\hat{t_0}_{\mbox {\tiny MME}}) &=&\mbox{Var}(\hat{t_0}_{\mbox {\tiny MME}})\nonumber\\
& =& \mbox{Var}(\delta_0)+\mbox{Var}(\hat{\mu}_{\mbox {\tiny MME}})  \nonumber\\
& =& \textstyle \frac{1-p}{p^2} \cdot \frac {n}{n-1},
\end{eqnarray}
based on Equation (\ref{equ:Var_MME_u}) and
$\mbox{Var}(\delta_0)=\frac{1-p}{p^2}$. Note that when $n \gg 1$,
$\mbox{MSE}(\hat{t_0}_{\mbox {\tiny MME}}) \approx \textstyle
\frac{1-p}{p^2}$.

\subsection{Linear Regression Estimator}

Since $\hat{t_0}_{\mbox {\tiny LRE}}=t_0+\delta_0-\hat{\mu}_{\mbox
{\tiny LRE}}$ and $\mbox{E}(\delta_0)=\mbox{E}(\hat{\mu}_{\mbox
{\tiny LRE}})=\mu$,
\begin{equation}
    \mbox{Bias}(\hat{t_0}_{\mbox {\tiny LRE}}) =
    t_0+\mbox{E}(\delta_0)-\mbox{E}(\hat{\mu}_{\mbox {\tiny
    LRE}})-t_0 = 0
\end{equation}
\begin{equation}
    \mbox{MSE}(\hat{t_0}_{\mbox {\tiny LRE}})
    =\mbox{Var}(\hat{t_0}_{\mbox {\tiny LRE}}) =
    \mbox{Var}(\delta_0-\hat{\mu}_{\mbox {\tiny LRE}}).
\end{equation}

\noindent Note that from Equations (\ref{equ:1}) and (\ref{equ:2}),
$\hat{\mu}_{\mbox {\tiny
LRE}}=\frac{12}{n^2-1}\sum_{i=1}^{n-1}{\frac{i(n-i)}{2n}\delta_i}$
that is independent of $\delta_0$. Hence,
\begin{eqnarray}
\mbox{MSE}(\hat{t_0}_{\mbox {\tiny LRE}})&=&\mbox{Var}(\hat{t_0}_{\mbox {\tiny LRE}})\nonumber  \\
&=& \mbox{Var}(\delta_0) + \mbox{Var} (\hat{\mu}_{\mbox {\tiny LRE}}) \nonumber \\
&=& \textstyle \frac{1-p}{p^2}
\cdot\frac{5n^3+6n^2-5n+6}{5n(n^2-1)},
\end{eqnarray}
based on Equation (\ref{equ:Var_LRE_u}) and
$\mbox{Var}(\delta_0)=\frac{1-p}{p^2}$. Note that when $n \gg 1$,
$\mbox{MSE}(\hat{t_0}_{\mbox {\tiny LRE}}) \approx \textstyle
\frac{1-p}{p^2}$.


\ifCLASSOPTIONcaptionsoff
  \newpage
\fi

\bibliography{./TDSC}

\end{document}